\documentclass[aps,prl,twocolumn,amsmath,amssymb,showpacs]{revtex4}
\usepackage{graphicx}

\newcommand{\e}{\epsilon}
\newcommand{\Ha}{\hat H}

\begin{document}
\title{Renormalization approach to non-Markovian open quantum system
  dynamics}
\author{Giulia Gualdi} 
\affiliation{Theoretische Physik, Universit\"{a}t Kassel,
  Heinrich-Plett-Str. 40, D-34132 Kassel, Germany} 
\author{Christiane P. Koch}
\affiliation{Theoretische Physik, Universit\"{a}t Kassel,
  Heinrich-Plett-Str. 40, D-34132 Kassel, Germany} 
\date{\today}
\begin{abstract} 
  We show that time induces a dynamical renormalization of the
  system-environment coupling in open quantum system dynamics. 
  The renormalizability condition,
  of the interactions being either local, or, alternatively,
  defined on a finite continuum support, is generally
  fulfilled for both discrete and continuous environments. 
  As a consequence, we find a Lieb-Robinson bound to hold
  for local and, surprisingly, also for non-local interactions. 
  This unified picture allows us to devise a controllable
  approximation for arbitrary non-Markovian 
  dynamics with an \textit{a priori} estimate of the
  worst case computational cost. 
\end{abstract}

\maketitle

\paragraph{Introduction}
The interaction with its environment
causes a quantum system to loose energy and phase --
this is termed decoherence~\cite{Breuer}. 
Decoherence poses a severe
challenge to the application of quantum technologies 
since a quantum system can never completely be isolated from its
environment. 
On the other hand, the effect of the environment is not necessarily
detrimental and can likewise be used for robust implementation of quantum
processes~\cite{BarreiroNat11}. 
A rigorous treatment of decoherence is challenging because
the system and environmental degrees of
freedom become entangled due to their interaction.
This entanglement is neglected when
invoking the Markov approximation~\cite{Breuer}. However, 
non-Markovian effects are abundant, in particular in the condensed
phase. Examples  of current interest include mechanical
oscillators close to their ground state~\cite{GenesAdvAMOP09},
carbon nanotubes and graphene~\cite{EichlerNatNano11},  
excitons of the light-harvesting complex
~\cite{SarovarNatPhys10}, and solid state devices
based on quantum dots~\cite{deLangeSci10}, superconducting
junctions~\cite{BluhmNatPhys10}, or nitrogen vacancy centers in
diamond~\cite{KolkowitzPRL12,TaminiauPRL12}.
A correct treatment of the non-Markovian dynamics is important 
for application of these systems in quantum technologies which require
a sufficient amount of control. Controllability is expected to be
better for non-Markovian than  Markovian systems due to the 
equivalence of  non-Markovianity and information backflow from the
environment to the system~\cite{BreuerPRL09}. Currently, non-Markovian 
dynamics are tackled with stochastic methods provided certain 
assumptions can be
made~\cite{PiiloPRL08,BreuerPRL08,KochPRL08}. Alternatively, 
one can simulate the non-Markovian dynamics for finite times,  
starting from a microscopic model for system and
environment and truncating the number of environmental degrees of freedom. 
This approach is particularly interesting in view of quantum devices
which always operate in finite time. It 
has been successfully employed in the context of
semiconductor quantum dots~\cite{deLangeSci10}, hydrogen
diffusion~\cite{BaerJCP97} and femtosecond photochemistry at
surfaces~\cite{KochJCP02,KochPRL03}, spin dynamics in 
NMR~\cite{HogbenJCP10} and the spin-boson toy
model~\cite{HughesJCP09}. The observation that 
the truncation approach works for such diverse systems
suggests an underlying general property of quantum dynamics:
Apparantly it takes time to establish correlations between system and
environment. Proving such a conjecture would allow one to rigorously
quantify the necessary ingedredients for an accurate and and efficient
simulation of open quantum systems.

Here, we  prove this conjecture and show that it yields 
a general approach to model decoherence. 
Our proof allows us to answer the question why a comparatively small
number of environmental degrees of freedom often turns out to be sufficient%
~\cite{deLangeSci10,BaerJCP97,KochJCP02,KochPRL03,HogbenJCP10,HughesJCP09}. 
Moreover, we show that no specific structure
of the environment and system-environment
interaction is required for the truncation approach to be applicable. 
This allows for simulating arbitrary open quantum system
dynamics with a prespecified accuracy, 
employing only a finite number of environmental modes. 

Our proof is  based upon a fresh look on decoherence by 
combining the Lieb-Robinson bound known in 
many-body physics~\cite{LiebCMP72,NachtergaeleNewTrends07} with 
the surrogate Hamiltonian method developed in physical
chemistry~\cite{BaerJCP97,KochJCP02,KochPRL03}.  Technically, 
for discrete environments, the Lieb-Robinson bound translates
the inherent locality of quantum dynamics into a quantitative estimate
for the information propagation speed.  We show that the notion of an
effective light cone can be used to set up a dynamical renormalization
procedure for the generator of the 'surrogate' evolution. For
continuous environments, the interactions are generally non-local. We
show that the concept of {\it quasi-finite resolution} represents the  
equivalent of quasi-locality for discrete environments. In both cases,
time naturally induces a renormalization of the system-environment
interaction. 

\paragraph{Discrete environments: quasi-locality of quantum dynamics}
We first consider the environment to be comprised of discrete degrees
of freedom.
Assuming the interactions between system and environment to be
bounded and, for simplicity,  bilinear, 
the total Hamiltonian can be defined on a generic lattice in 
arbitrary dimensions,
\begin{equation}
  \hat H=\hat H_S+\sum_{i=1}^{N^{int}_S}\sum_{j=1}^{N^{int}_B}\hat
  \Phi^{SB}_{ij}+\sum_{i\leq j=1}^{N_B}\hat\Phi^B_{ij} 
  \label{eq:gen2} 
\end{equation}
with $N_S$ and $N_B$ the system and environmental degrees of freedom
(DOF), $N_B\to\infty$. In Eq.~\eqref{eq:gen2}, 
each DOF is defined on a finite-dimensional Hilbert
space, $\mathcal{H}_i$. $N^{int}_{S}\leq N_S$ ($N^{int}_B\leq N_B$)
are those system (environmental) DOF that interact
with the environment (system). The interactions 
$\hat\Phi_{ij}$ can be expressed in terms of linear operators
$\hat O_i \in \mathcal{B}(\mathcal{H}_i)$,
\begin{displaymath}
  \hat\Phi_{ij}=\sum_{\mu=0}^{dim(\mathcal{B}(\mathcal H_i))-1}\;
  \sum_{\nu=0}^{dim(\mathcal{B}(\mathcal{H}_j))-1}J^{\mu\nu}_{ij}
  \hat O^{\mu}_i\hat O^{\nu}_j\,,
\end{displaymath}
with $|J^{\mu\nu}_{ij}|\!\!<\!\!\infty$. Our goal is to truncate the
sums over the environmental DOF in Eq.~\eqref{eq:gen2} in a
well-defined manner. To this end,
we need to quantify the influence of the DOF upon each other, i.e., we
need to introduce a metric. A suitable metric arises naturally
by representing the Hamiltonian $\hat H$ as a graph $G$.
The set of nodes of the graph is composed of all the 
system and environmental DOF, $N=\{N_S+N_B\}$, i.e., a possibly 
infinite number of elements. The edges of the
graph are made up by all non-zero elements of the coupling matrix 
$J_{ij}=\left[\sum_{\mu\nu}(J^{\mu\nu}_{ij})^2\right]^{1/2}$,
$E=\{J_{ij}\neq0\}$. 
The bare structure of the graph is encoded in the adjacency matrix
$A=A(G)$ whose entries $A_{ij}$ ($A_{ij}=0,1$) represent the edges
connecting two nodes $i$ and $j$.  The metric induced by $\hat H$
on $G$ is defined as the shortest path
connecting two nodes, 
\begin{displaymath}
  d(i,j):=\min\{n\in\mathbb{N}_0: [A^n]_{ij}\neq0\}\,.
\end{displaymath}
A walk of length $n$ from node $i$ to $j$ is a sequence
of $n$ adjacent nodes. Their weight is
$\prod_{k=1}^{n-1}{J}_{i_k,i_{k+1}}$ with the weight of the zero-length
walk set equal to 1. Then the overall weight of  
all paths of length $n$ between $i$ and $j$ is $[J^n]_{ij}$, and 
the weight of the shortest path(s) is $[{J}^{d(i,j)}]_{ij}$. 
Equipped with the metric $d(i,j)$, we can order the 
environmental DOF according to their graph 
distance from the system, i.e., their minimum distance 
from a node in $\{N_S^{int}\}$. For the sake of simplicity, we
consider a single system node $S$; the generalization to $N_S>1$ is 
straightforward. Reordering the environmental DOF is expressed by
rewriting Hamiltonian~\eqref{eq:gen2},
\begin{equation}
\hat H = \sum_{d=0}^{\infty}\left(\hat{h}_{d}+ \hat{h}_{d,d+1}\right)\,,
\label{eq:onion}
\end{equation}
where $\hat h_d$ groups the interactions  between DOF at distance 
$d$ from the system , i.e., those in the $d$th layer.
$\hat h_{d,d+1}$ contains the interactions between DOF 
in two successive layers, e.g.,
$\hat H_S=\hat h_0$ and $\hat H_{SB}=\hat h_{0,1}$.

The dynamical evolution of a  generic 
system operator $\hat A_S\in \mathcal B(\mathcal H_S)$ is given by 
$\hat A_S(t) = e^{i\Ha t}\hat{A}_S e^{-i\Ha t}$.
Using the Baker-Campbell-Hausdorff formula, $\hat A_S(t)$
can be written in terms of nested commutators, 
\begin{equation}
\hat A_S(t) = \hat A_S + \sum_{d=1}^{\infty}\frac{(-it)^d}{d!}\hat C_d\,,
\label{eq:BCH}
\end{equation}
where $\hat C_d=[\Ha,\hat C_{d-1}]$ and $\hat C_0=\hat A_S$. 
We show in the Supplementary Material~\cite{SM} that $\hat C_d$
has non-vanishing commutators only with those terms in
$\hat H$  that act on DOF in the layers $d'\leq d$, including $\hat 
h_{d,d+1}$. This implies 
$\hat C_n=[\hat H, \hat C_{n-1}]\equiv[\hat H_{n},\hat C_{n-1}]$
at the $n$th perturbative order, where 
\begin{equation}
\Ha_n=\sum_{d=0}^{n-1}\left(\hat h_{d}+ \hat h_{d,d+1}\right)
\label{eq:onion2}
\end{equation}
is the truncation of the full generator $\hat H$ to the first $n$
layers of the graph. 
In other words, terms corresponding to bath DOF at distance $n$ from
the system start contributing to the system dynamics only at the $n$th
perturbative order. The system dynamics is thus appreciably affected
by those bath modes only after a time that is sufficiently long to make
the corresponding perturbative term non-negligible. 

In order to make this statement quantitative, we
consider the error made by truncating the full evolution, $\hat
A_S(t)$, at the  $n$th perturbative order, 
$\hat A_S^{n}(t)$, i.e., the error made by replacing $\hat H$  
by the truncated generator~\eqref{eq:onion2},
$\mathcal{R}(n)=\left\|\hat A_S(t)-\hat A_S^{n}(t)\right\|$. This error
corresponds to the remainder of the series in Eq.~\eqref{eq:BCH} when
it is truncated at order $n$~\cite{SM},
\begin{displaymath}
\mathcal{R}(n) \,\leq\, \left\|\hat A_S\right\|
\sum_{d=n+1}^\infty\frac{\left(2t\mathcal{O}\right)^d}{d!}
\sum_{i,j\in\mathcal{I}_d}[J^d]_{ij}\,,
\end{displaymath}
where $\mathcal O=\max_{(i,j)\in N; \mu,\nu}\left\|\hat O^\mu_i\hat
O^\nu_j\right\|$, and 
$\mathcal{I}_d=\{i \in N: d(s,i)\leq d\}$ represents the set of DOF at
distance  at most $d$ from the system. $\sum_{i,j
\in\mathcal{I}_d}[J^d]_{ij}$ is the weight  of all paths of length $d$
involving  DOF at distance at most $d$ from the system. 
If the graph is \textit{locally finite}, i.e., if each DOF interacts
with  a finite number of other DOF, then
$\|J\|<\infty$~\cite{MoharBLMS89}. In this case, we can bound the sum,  
$\sum_{i,j\in\mathcal{I}_d}[J^d]_{ij}\leq \left(\bar c^2\|J\|\right)^d$, where 
$\bar c$ denotes the maximum connectivity of a node on the graph.  
This leads to the following Lieb-Robinson
bound~\cite{LiebCMP72,NachtergaeleNewTrends07,BravyiPRL06,OsbornePRL06}
\begin{equation}
\left\|\hat A_S(t)-\hat A_S^{n}(t)\right\|\leq 
\left\|\hat A_S\right\|e^{-(n-vt)}
\label{eq:LR}
\end{equation}
with  $v=2\mathcal{O}\bar c^2\|J\|e$~\cite{SM}. 
Equation~\eqref{eq:LR} states quasi-locality of the
quantum  dynamics of an open quantum system. 
It implies the existence of an effective 
light-cone for the system that spreads at most at speed $v$. 
Bath DOF outside of the effective light cone give only an
exponentially vanishing contribution to the evolution of $\hat A_S$.
The full bath is therefore needed only in the limit of infinite time.

\paragraph{Continuous environments: 
quasi-finite resolution of quantum dynamics}
We now consider the interaction of a central system with 
a continuous environment. The  corresponding Hamiltonian  is
generically expressed by
\begin{eqnarray}
  \hat H&=&\hat H_S+
  \hat O^I_S\int_{0}^{x_{max}}J(x)\left(\hat c_x+\hat
    c_x^\dagger\right)dx \nonumber\\
  && + 2\int_0^{x_{max}}\!\!\int_{x}^{x_{max}}
  K(|x-x'|)\bigg[ c_xc^\dagger_{x'}+ c^\dagger_xc_{x'} \nonumber\\
  && + c^\dagger_x\hat c_xc^\dagger_{x'}\hat c_{x'}\bigg] dx\,dx' 
  +\int_{0}^{x_{max}}g(x)\hat c_x\hat c_x^\dagger dx\,,
\label{eq:cont} 
\end{eqnarray}
where $x$ denotes the relevant bath variable such as energy or
position, $x_{max}<\infty$ is a finite cut-off, and
$\hat O^I_S$ a generic system operator. We require the 
annihilation (creation) operators of bath modes, 
$\hat c_x(\hat c^\dagger_x)$, to be bounded, i.e., 
$\|c\|=\max_{x\in[0,x_{max}]}\|\hat c_x\|<\infty$. $J(x)$ denotes the
system-bath coupling, $K(|x-x'|)$ the intra-bath coupling,  assumed 
symmetric under exchange of $x$ and $x'$, and $g(x)$
is the bath dispersion. 
The Hamiltonian~\eqref{eq:cont} does not obey local finiteness,
since the system interacts
with all bath DOF which, in turn, all may interact among themselves.
This corresponds to a graph where all bath DOF are at distance $1$  
from the system, such that the results of the previous
section cannot be used directly to truncate the Hamiltonian. If the
bath is made up of normal 
modes, $K(|x-x'|)\equiv0$, then~\eqref{eq:cont} can be mapped onto a 
semi-infinite chain  with the system at one
end~\cite{PriorPRL10,ChinJMP10}, thus recovering local finiteness. 
However, this requires the bath Hamiltonian to be quadratic and is thus not
general. Here, we derive a generally applicable bound equivalent to
Eq.~\eqref{eq:LR} for continuous environments, employing the concept
of a 'surrogate Hamiltonian'~\cite{BaerJCP97}. 

We choose a sequence of $n$ points $\{x_i\}_{i=0}^{n-1}$,
in the interval $[0,x_{max}]$, with $x_i<x_{i+1}$, 
thus defining a partition $P_n=\{\delta x_i\}$ of the interval with
$\delta x_i=x_{i+1}-x_i$. Denoting the norm of $P_n$ by  
$|P_n|=\max_{i<n}(\delta x_i)$, a sequence of
partitions $\{P_n\}$ obeying the condition $|P_{n+1}|<|P_n|$ 
can be constructed. This defines a sequence  of Hamiltonians
$\{\hat H_{P_n}\}$  with
\begin{eqnarray}
\hat{H}_{P_n} &=&\hat H_S+\hat O^I_S\sum_{i=0}^{n-1}\tilde J_i
(\hat c_i+\hat{c}^\dagger_i)+\nonumber\\
&&2\!\sum_{i<j=0}^{n-1}\!\tilde{K}_{ij}\!\left[
\hat c^\dagger_i\hat c_j+\hat c_j^\dagger\hat c_i+
\hat c^\dagger_i\hat c_i\hat c^\dagger_j\hat c_j\right]\nonumber \\ &&+\!
\sum_{i=0}^{n-1}\tilde{g}_i \hat c^\dagger_i\hat{c}_i,
\label{eq:N}\end{eqnarray}
where $\hat{c}_i=\hat c_{x_i}$,
$\tilde{J}_i=J(x_i)\delta x_i$, $\tilde K_{ij}=K(|x_i-x_j|)\delta
x_i\delta x_j$, and $\tilde{g}_i=g(x_i)\delta x_i$ are the rescaled
couplings at the $n$ sampling
points. Equation~\eqref{eq:N} can be viewed as  Riemann sums
built on $P_n$ approximating the corresponding integrals in
Eq.~\eqref{eq:cont}. By construction, the sequence $\{\hat H_{P_n}\}$ 
converges with $\lim_{n\rightarrow\infty}\hat H_{P_n}=\hat H$. 

When estimating the error that is 
made by time evolving $\hat A_S$ using $\hat H_{P_n}$ instead of
$\hat H$, $\mathcal{R}(P_n)=\|\hat A_S(t) - \hat A_S^{H_{P_n}}(t)\|$, 
we need to compare two series of the kind~\eqref{eq:BCH}, 
one for $\hat A_S(t)$ and one for $\hat A_S^{H_{P_n}}(t)$.
The triangle inequality can be used to
split the error into two parts,
$\mathcal{R}(P_n)\leq R_1(P_n) + R_2(P_n)$~\cite{SM}.
The first term evaluates the error made in replacing
$\exp[-i\hat Ht]\exp[i\hat H_{P_n}t]$ by 
$\exp[-i(\hat H-\hat H_{P_n})t]$, i.e.,
assuming  $\hat H$ and $\hat H_{P_n}$ to
commute. This contribution
is bounded by a second order polynomial  in 
$t^2\left\|[\hat H,\hat H_{P_n}]\right\|$. At finite times
$R_1(P_n)$ vanishes in the limit $n\rightarrow\infty$ due to the
convergence of Riemann sums. 
The second contribution to the error, $R_2(P_n)$ represents the
distance between $\hat A_S$ and its evolution under $\hat H-\hat
H_{P_n}$.
As final estimate we obtain
\begin{eqnarray}
  \label{eq:finalbound}
  \left\|\hat A_S(t) - \hat A_S^{H_{P_n}}(t)\right\| &\leq&
  R_1(P_n) \\ \nonumber
  &&+\left\|\hat A_S\right\|\left(e^{2\left\|\hat H-\hat H_{P_n}\right\|t}-1\right)\,,
\end{eqnarray}
with $R_1(P_n)$ given in Eq.~(34) of the Supplementary Material~\cite{SM}.
Equation~\eqref{eq:finalbound} states \textit{quasi-finite resolution}
of quantum dynamics: At finite times one can 
reproduce the system-bath dynamics within arbitrary accuracy by 
employing an effective generator, Eq.~\eqref{eq:N}, that is 
constructed on a finite number of sampling points with rescaled
couplings.  The full continuum of environmental modes is resolved 
only in the limit of  infinite time. 
Given a specific form for the couplings in the
Hamiltonian~\eqref{eq:cont}, a clever choice of the sampling  
can prolong convergence times. 

Equations~\eqref{eq:LR} and \eqref{eq:finalbound} provide an upper
bound to the error made by replacing the full generator, $\hat H$, by an
effective one, $\hat H_d$ or $\hat H_{P_n}$.  
The bounds are general. They are therefore also very
conservative. In some specific cases, tighter model-dependent bounds can be
derived~\cite{BurrellPRL07,HastingsPRB08}. For certain classes of initial
states, the scaling with time can be dramatically
reduced~\cite{HastingsPRB08,EisertRMP10}. 
Extension of the bounds, Eqs.~\eqref{eq:LR} and \eqref{eq:finalbound},
to $k$-linear interactions is straightforward~\cite{SM}.
However, extension to Hamiltonians containing unbounded operators,
i.e., $\mathcal{O}=\infty$, is possible only for certain classes of
operators~\cite{Cramer08,NachtergaeleRMathP10}. 
\begin{figure*}[tb]
\hspace*{0.05\linewidth}\includegraphics[width=0.35\linewidth]{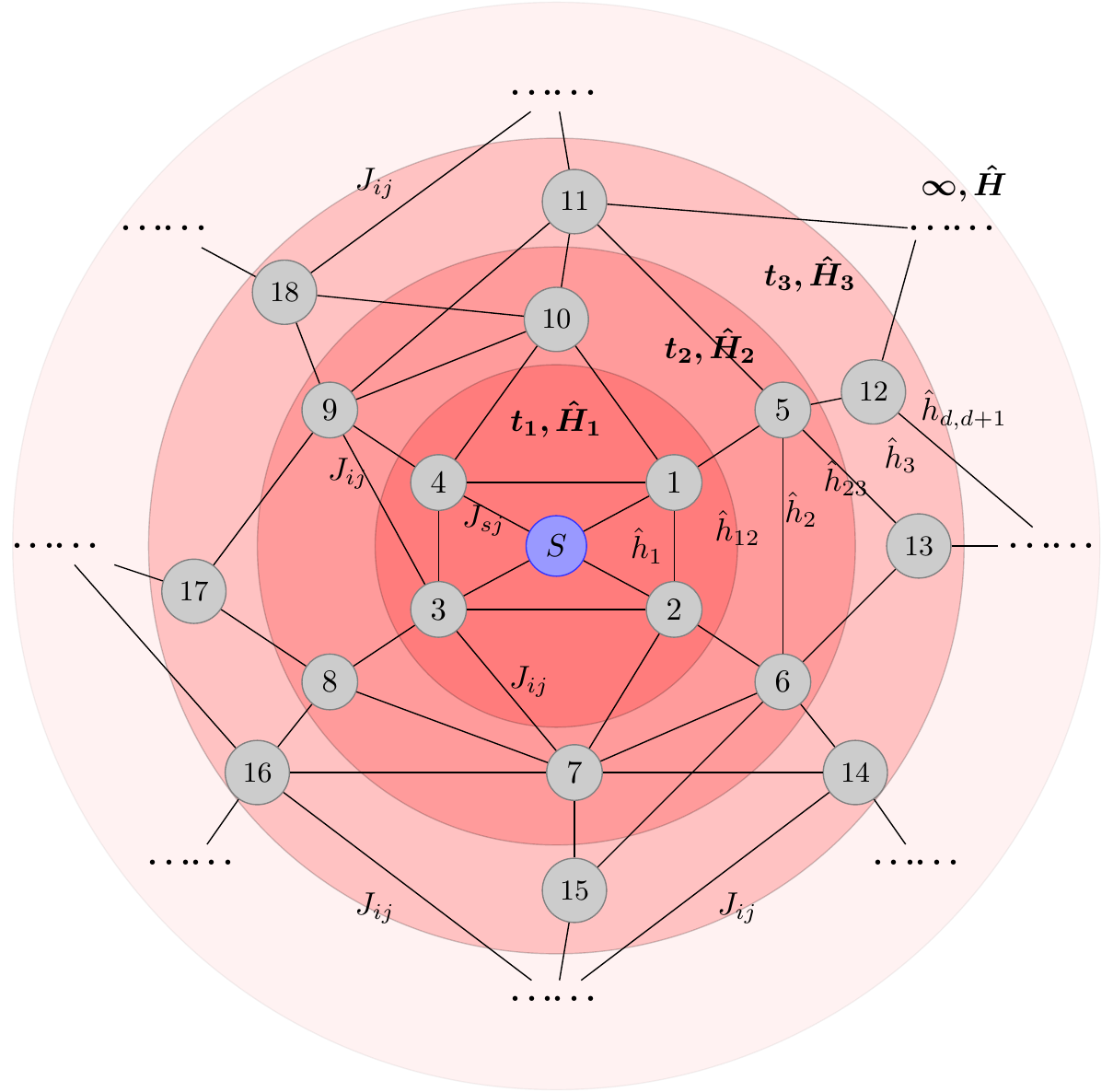}%
\hspace*{0.10\linewidth}%
\includegraphics[width=0.45\linewidth]{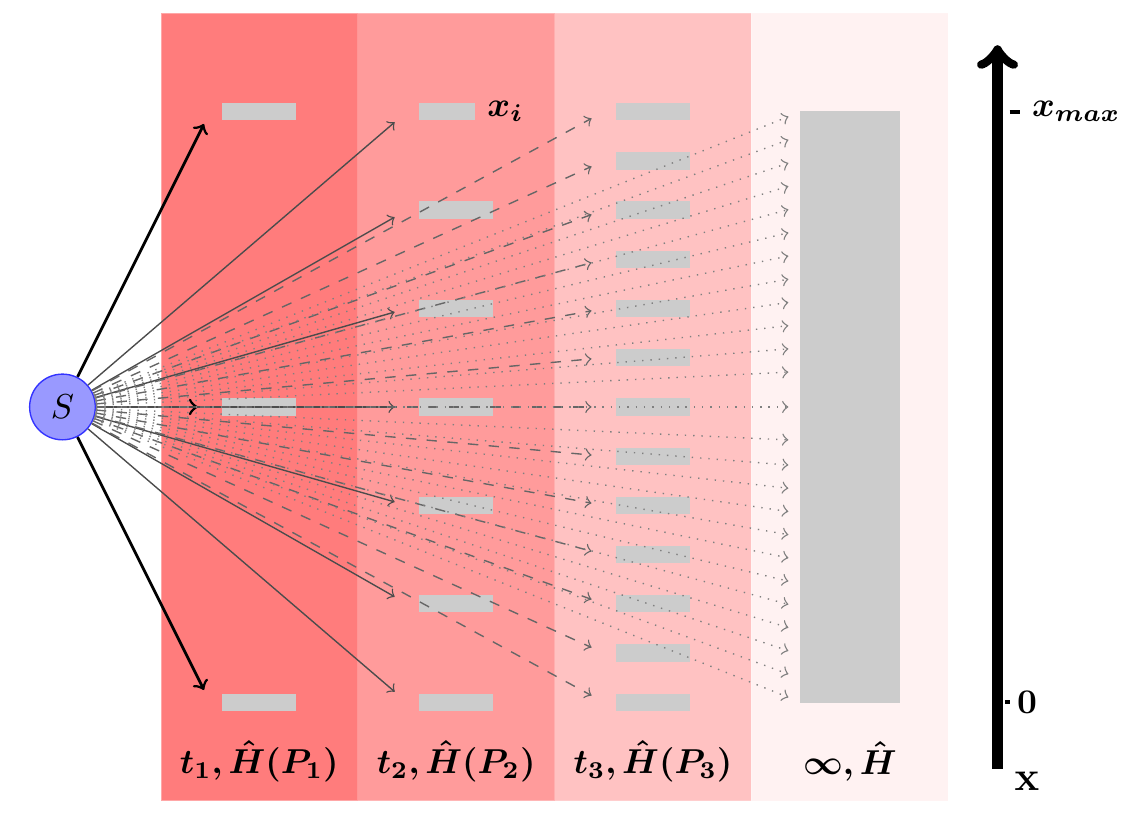}
\caption{\label{fig1}(color online) 
  Dynamical renormalization of the system-environment coupling.
  Discrete environments (left):
  The effective generator $\hat H_{d}$ is updated by adding the interaction
  with all environmental degrees of freedom in the $d+1$st layer and the new local
  terms. Continuous environments (right):
  The effective generator $\hat H_{P_n}$ is updated  to $\hat H_{P_{n+1}}$ 
  by adding sampling points in the interval $[0,x_{max}]$ and rescaling
  the couplings.  
 } 
\end{figure*}

\paragraph{Correspondence between discrete and continuous environments} 
The Hamiltonian for a system interacting with a continuous
environment, Eq.~\eqref{eq:cont}, corresponds to 
an infinite graph whose environmental nodes are all at distance $1$
from the system. 
The Hamiltonians~\eqref{eq:onion} and \eqref{eq:cont} thus represent
the two opposite extremes of an infinite graph -- with the infinite
number of environmental nodes concentrated in a single layer or
distributed over infinitely many layers.
In both cases, the system-bath coupling can be defined as the weight
$\mathcal{J}_{SB}$ 
of the paths needed by the system to explore all of the environment. 
For continuous environments,
$\mathcal{J}_{SB}=\int_0^{x_{max}}J(x)dx$, 
which is finite because the support of the integral is finite. 
For discrete environments,
$\mathcal{J}_{SB}=\sum_{n=0}^{\infty}\sum_{j:d(s,j)=n}[J^n]_{sj}$, and 
local finiteness ensures that $\mathcal{J}_{SB}$ can be made 
finite by rescaling the coupling matrix, e.g., by setting $\tilde
J=J/r$ with $r\geq \bar c\|J\|$. 
This amounts to penalizing longer paths and allows for bounding all 
quantities on the graph. The dynamics in Hilbert space remains
unaffected since any rescaling of the coupling matrix  is cancelled
out by a corresponding rescaling of time ($t\rightarrow\tilde t=rt$). 
Local finiteness and the finite cutoff $x_{max}$ 
thus play the same role for the two representations of infinitely
large environments, with infinitely long paths the discrete
counterpart of infinitely close modes.  

\paragraph{Dynamical renormalization}
This unified picture for discrete and continuous environments implies
that in both cases time naturally induces a dynamical renormalization
over the system-bath interaction. It is expressed by 
the bounds, Eqs.~\eqref{eq:LR} and \eqref{eq:finalbound}, which provide
a recursive update rule for the effective
generators, as illustrated in  Fig.~\ref{fig1}. 
For discete environments, the number of required bath modes is
obtained as function of time, $n=n(t)$, by 
specifying the desired accuracy and inverting  
Eq.~\eqref{eq:LR}. 
Defining $\tilde{\mathcal{J}}(n(t))=\sum_{d=0}^{n(t)}\sum_{j:d(s,j)=d}
\left[\tilde{J^d}\right]_{sj}$, the renormalization flow 
is expressed as
$\lim_{t\rightarrow\infty}\tilde{\mathcal{J}}(n(t))=
\tilde{\mathcal{J}}_{SB}$. 
For continuous environments, given a desired accuracy and simulation
time, the required resolution is obtained from
Eq.~\eqref{eq:finalbound} as $|P_n|=|P_{n(t)}|$.
The renormalization flow corresponds to the convergence of
Riemann sums, $\mathcal{J}(P_{n(t)})=\sum_{i\in P_{n(t)}}\tilde J_i$ as 
$\lim_{t\rightarrow\infty}\mathcal{J}(P_{n(t)})=
\mathcal{J}_{SB}$.

Due to Eqs.~\eqref{eq:LR} and \eqref{eq:finalbound} 
the dynamics of any open quantum system can  be  simulated
efficiently on a quantum computer: Once the 
generator is defined on a finite Hilbert space, it can be approximated
by a Suzuki-Trotter 
decomposition~\cite{Trotter59,SuzukiProc93,DeRaedtCPR87,PoulinPRL11}. This
represents  a quantum circuit, i.e.,  a sequence of  
elementary quantum gates for each time  step.
The cost of simulating the effective generator scales only
polynomially in time and the number of effective degrees of freedom~\cite{SM}.
The resources required on a classical computer are, however, exponential
in the number of effective environmental degrees of freedom. 
This is due to the exponential
scaling of the system plus bath state that needs to be stored.
It is in contrast to uncontrollable approximations such as the Markov
approximation where the environment is completely eliminated from the
reduced dynamics such that the computational resources are constant with time
and only depend on the size of the system Hilbert space.
The exponential scaling of the computational resources with the number
of effective degrees of freedom can be reduced to a polynomial one by employing further
controlled restrictions of the size of the effective Hilbert 
space~\cite{KochJCP02,KochPRL03,HogbenJCP10,VidalPRL04,WhitePRL04,DaleyJSM04}.

\paragraph{Conclusions} 

We have shown that the reduced dynamics of an arbitrary open quantum
system can be obtained reliably and accurately, 
employing only a finite-dimensional effective
Hamiltonian. This is due to time inducing a dynamical renormalization of 
the system-environment interaction, i.e.,  the system
interacts progressively with the environmental degrees of freedom rather than with
all of them at once. The required renormalizability condition, 
locality of the interactions for discrete environments and 
finite support of the interactions for continuous environments, is 
generally fulfilled. 
While the Lieb-Robinson bound has been discussed
in the context of dissipative dynamics
before~\cite{PoulinPRL10,HastingsPRL04}, it was never employed for 
the full system-bath evolution. Carrying out this very natural
application of the bound, we have generalized the notion of  
quasi-locality of quantum dynamics to non-local interactions. In spin
dynamics, the Lieb-Robinson bound provides the theoretical foundation
of truncation-based algorithms such as
t-DMRG~\cite{OsbornePRL06}. Similarly, 
our results allow to assess the worst case
computational cost of  truncation-based algorithms for
non-Markovian  dynamics  and certify
\textit{a priori} their accuracy versus computational
complexity.

\begin{acknowledgments}
  We would like to thank M. Cramer, S. Gualdi, R. Kosloff, K. Murr,
  and D. M. Reich for valuable discussions. Financial support from the
  EC through the FP7-People IEF Marie Curie action QOC4QIP Grant
  No. PIEF-GA-2009-254174 is gratefully acknowledged.
\end{acknowledgments}



\begin{widetext}
\begin{center}
{\large Supplementary Material:\\
Renormalization approach to non-Markovian open quantum system
  dynamics}

\bigskip
Giulia Gualdi and Christiane P. Koch

\medskip
\textit{Theoretische Physik, Universit\"{a}t Kassel,
Heinrich-Plett-Str. 40, D-34132 Kassel, Germany}
\end{center}  
\end{widetext}

\bigskip
\section{Quasi-locality of quantum dynamics for discrete 
  environments} 

In this section we prove the quasi-locality of quantum dynamics
for a system interacting with an environment comprised of discrete
degrees of freedom (DOF). In this case, the total system+environment
is  defined on a lattice.
The starting point is the Hamiltonian, Eq.~(3) in the main text, 
written in terms of 
$\hat h_d$, grouping the interactions between DOF at graph distance 
$d$ from the system region, i.e.,  in the $d$th layer, and $\hat
h_{d,d+1}$, comprising the interactions between DOF in two successive
layers.  With these definitions, 
$\hat H_S\equiv\hat h_0$ and $\hat H_{SB}\equiv\hat h_{0,1}$.
If the system  is made up of more than a single node, the graph
distance is calculated as the minimum between a
bath and a system DOF.  

The dynamics of a generic system operator, $\hat A_S\in \mathcal
B(\mathcal H_S)$, is described by
\[
\hat A_S(t) = e^{i\Ha t}\hat{A}_Se^{-i\Ha t}\,.
\]
Using the Baker-Campbell-Hausdorff formula, this can be expressed in
terms of nested commutators, cf. Eq.~(4) of the main paper,
where the commutator $\hat C_n$, 
\[
\hat C_n=[\Ha,\hat C_n-1]\,,
\] 
appears at the $n$th
order, and $\hat C_0=\hat A_S$. The truncation of the sum in Eq.~(4)
of the main paper can be identified, to the
$n$th order in a perturbative treatment, by restricting 
the Hamiltonian to $n$ graph layers using the graph-distance based
ordering of the DOF, cf. Eq.~(5) of the main paper.
The error made by evolving $\hat A_S$ with $\hat H_n$ instead of 
the full generator $\hat H$ can be evaluated as the remainder of the 
truncated series, 
\begin{equation}
  \left\|\hat A_S(t)-\hat A_S^{n}(t)\right\| =\left \|\sum_{d=n+1}^\infty\ 
    \frac{(-it)^d}{d!}\hat C_d\right\|\equiv \mathcal{R}(n)\,,
  \label{eq:err}
\end{equation}
where $\hat A^n_S(t)=\exp(i\hat H_nt)\hat A_S\exp(-i\hat H_nt)$.
The triangular inequality yields 
\begin{equation}
  \mathcal{R}(n) \leq
  \sum_{n+1}^{\infty}\frac{t^d}{d!}\left\|\hat C_d\right\|.
  \label{eq:Rn}
\end{equation}
In order to estimate $\left\|\hat C_d\right\|$, we need to consider
the following commutators between operators in $\hat H$,
\begin{widetext}
\begin{eqnarray}
  \left[\hat h_d\;,\;\hat h_{d'}+\hat  h_{d',d'+1}\right]
  &=& \left[\hat h_d\;,\;\hat  h_{d,d+1}\right]\delta_{d,d'} 
  +\left[\hat h_d\;,\;\hat  h_{d-1,d}\right]\delta_{d-1,d'}\, \label{eq:loc}\\
  \left[ \hat {h}_{d,d+1}\;,\;\hat h_{d'}+\hat  h_{d',d'+1}\right]
  &=&\left[\hat h_{d,d+1}\;,\;\hat h_{d}\right]\delta_{d,d'}
  +\left[\hat h_{d,d+1} \;,\;
    \hat h_{d+1}+\hat h_{d+1,d+2}\right]\delta_{d+1,d'}
  +\left[\hat h_{d,d+1},\hat  h_{d-1,d}\right]\delta_{d-1,d'}\,.
  \label{eq:nonloc}
\end{eqnarray}  
\end{widetext}
'Local' operators $\hat h_d$, 
i.e., terms involving interactions between DOF
within the same layer, have non-vanishing commutators
only with operators connecting the $d$th layer with the neighbouring
layers, $d\pm 1$, cf. Eq.~\eqref{eq:loc}.
Since the $(d+1)$st layer is already accounted for in $\hat
h_{d,d+1}$, commutators of 'local' terms $\hat h_d$,
Eq.~\eqref{eq:loc}, 
do not introduce further layers. In other words, these
commutators do not increase the size of
the bath Hilbert space that is 'seen' by the system. 
Terms involving interactions between the $d$th and the $(d+1)$st
layer, i.e., $\hat h_{d,d+1}$, have non-vanishing commutators at most
with terms in the same, the previous or the $(d+2)$nd layer. 
Therefore, commutators involving the non-local terms, $\hat
h_{d,d+1}$, add operators from one additional layer, i.e., they
enlarge the system 'view' by one graph layer at each 
perturbative order.    
Examining the generic structure of the $\hat C_d$'s, we find for 
$\hat C_1$, 
\begin{displaymath}
  \hat C_1=\left[\hat H\;,\;\hat A_S\right]
  \equiv\left[\hat h_0+\hat h_{0,1}\;,\;\hat A_S\right].
\end{displaymath}
We can rewrite
\begin{displaymath}
\hat h_0+\hat h_{0,1}=\sum_{\substack{i,j:\\
d(S,i)=0;\\ d(S,j)\leq 1}}
\sum_{\mu,\nu}J^{\mu\nu}_{ij}\hat O^{\mu}_i\hat O^\nu_j\,,
\end{displaymath}
using Eq.~(2) of the main text and the definition of 
$\hat\Phi_{ij}$'s. This 
highlights the fact that $\hat h_0+\hat h_{0,1}$ groups all
interactions within the system and between system and bath. It
implies 
\begin{eqnarray}
  \hat C_1&=&\sum_{\substack{i,j:\\
      d(S,i)=0;\\ d(S,j)\leq 1}}
  \sum_{\mu,\nu}J^{\mu\nu}_{ij}
  \left[\hat O^{\mu}_i\hat O^\nu_j\;,\;\hat A_S\right]\nonumber\\
  &=&\sum_{j} \sum_{\mu,\nu}J^{\mu\nu}_{Sj}
    \left[\hat O^{\mu}_S\hat O^\nu_j\;,\;\hat A_S\right]
\label{eq:C1}\end{eqnarray}
In the second line, we have used that only commutators between
operators which act at least on one common DOF  do not vanish, 
assuming, for the sake of clarity, that $\hat A_S$ acts on a single
system DOF. Should $\hat A_S$ act on multiple system DOF, 
a sum over these DOF needs to be included additionally. 
Introducing 
\[
\mathcal{O}= \max_{ij\in N;\mu,\nu}\|\hat O^\mu_i\hat O^\nu_j\|\,,
\]
the norm of $\hat C_1$ can be estimated, 
\begin{displaymath}
  \|\hat C_1\|\leq 2\|A_S\|\mathcal{O}\sum_{j}J_{Sj}\,,
\end{displaymath}
where $J_{ij}=\left[\sum_{\mu\nu}(J^{\mu\nu}_{ij})^2\right]^{1/2}$ denotes
the coupling matrix on the graph.
In the following, we drop the indices $\mu,\nu$, accounting for the
corresponding sums in the coupling matrix $J$, and 
denote the system operator that enters the system-bath interaction by
$\hat O^I_S$. 
Due to Eqs.~\eqref{eq:loc}, \eqref{eq:nonloc}, $\hat C_2$ is written
as 
\[
  \hat C_2=[\hat h_0+\hat h_{0,1}+\hat h_1+\hat h_{1,2},\hat C_1]\,.
\]
Using Eq.~\eqref{eq:C1} and following the same argument we find
\begin{eqnarray}
  \hat C_2
  &=&\sum_{\substack{p,q:\\d(S,p)\leq 1\\ d(S,q)\leq 2}}
  \!\sum_{j:d(S,j)\leq 1}
  J_{pq}J_{Sj}\left[\hat O_p\hat O_q\;,\;
    \left[\hat O^I_S\hat O_j\;,\;\hat A_S\right]\right]\nonumber\\
  &=&\sum_{\substack{j,p:\\d(S,j)\leq 1\\d(S,p)\leq2}}
  J_{pj}J_{Sj}\left[\hat O_p\hat O_j\;,\;
    \left[\hat O^I_S\hat O_j\;,\;\hat A_S\right]\right].\label{term1}
\end{eqnarray}
So $\hat C_2$ 
groups the commutators along paths of length
$2$ that either depart from the system without returning to it ($j\neq S,p\neq S$),
or that pass through it one or two times ($j=S$ and/or $p=S$).
Analogously to the norm of $C_1$, we can estimate $\left\|\hat
  C_2\right\|$, 
\begin{displaymath}
  \left\|\hat C_2\right\| \leq 
  \left\|A_S\right\|\left(2\mathcal{O}\right)^2
  \sum_{i,j\in \mathcal I_2}\left[J^2\right]_{ji}\,,
\end{displaymath}
where $\mathcal I_2=\{i\in N: d(S,i)\leq 2 \}$ 
denotes the set of nodes at distance at most $2$ from the system, and   
$\sum_{i,j\in \mathcal I_2}[J^2]_{ji}$ is the weight of all paths
of length $2$ that exist between DOF in $\mathcal {I}_2$. 
Iterating this procedure, the  general form of the $\hat C_d$'s is
found to be 
\begin{widetext}
\begin{eqnarray}
  \label{An}
  \hat C_d=\sum_{\substack{(i_1,j_1)\in\mathcal{I}_1\\\vdots\\(i_d,j_d\in\mathcal{I}_d)}}
  \,\prod_{k\in[1,d]}J_{i_k,j_k} 
  \left[\hat O_{i_d}\hat O_{j_d}\;,\;
    \left[\hat O_{i_{d-1}}\hat O_{j_{d-1}}\;,\;
      \left[\cdots\left[\hat O_{i_1}\hat O_{j_1}\;,\;
          \hat A_S\right]\cdots
      \right]
    \right]
  \right]\,.
\end{eqnarray}  
\end{widetext}
The set $\mathcal{I}_k=\{i\in N: d(s,i)\leq k\}$ contains the DOF
belonging to  the first $k$ layers of the graph. Thus
$\mathcal{I}_k\subseteq\mathcal{I}_{k+1}$.
Due to the presence of the commutators and Eqs.~\eqref{eq:loc},
\eqref{eq:nonloc}, the only non-vanishing terms 
in the sums over the $\mathcal{I}_k$ are those where all adjacent
pairs of indices have at least one element in common, 
i.e., those of the general form
\begin{displaymath}
  \sum_{\alpha_{d-1}\in\mathcal{I}_{d-1}}\sum_{\alpha_{d-2}\in\mathcal{I}_{d-2}}
  \cdots\sum_{\alpha_0\in\mathcal{I}_0}J_{i,\alpha_{d-1}}\prod_{k=0,d-2}J_{\alpha_{k+1},\alpha_k}
\end{displaymath}
with $i\in\mathcal{I}_d$. 
Each of these terms represents a path of length $d$ 
within the first $d$ layers of the graph.
We can therefore estimate 
\begin{equation}
  \left\|\hat C_d\right\|\leq 
  \left\|\hat A_S\right\|\left(2\mathcal{O}\right)^d
  \sum_{i,j\in\mathcal{I}_d}\left[J^d\right]_{ij}\,,
\label{max}\end{equation}
where the sum accounts for all paths of 
length $d$ between two DOF that are at most at distance $d$ from the
system. Hence, we can rewrite the remainder, Eq.~\eqref{eq:Rn},
\begin{displaymath}
  \mathcal{R}(n) \leq 
  \left\|\hat A_S\right\|\sum_{d=n+1}^\infty
  \frac{\left(2t\mathcal{O}\right)^d}{d!}
  \sum_{i,j\in\mathcal{I}_d}\left[J^d\right]_{ij}\,.
\end{displaymath}
If the graph is \textit{locally finite}, i.e., if each DOF is
connected to a finite amount of other DOF, then 
\begin{equation}
  \label{eq:lf1}
  \|J\|\leq\max_{i\in  N}\sum_jJ_{ij}\leq\infty\,.   
\end{equation}
Labeling the maximum  
connectivity of a node on the graph by $\bar c$, we can estimate 
\begin{equation}
  \label{eq:lf2}
  \sum_{i,j\in\mathcal{I}_d}\left[J^d\right]_{ij}\leq 
  \left(\bar c^2\left\|J\right\|\right)^d\,,  
\end{equation}
since the relevant part of the coupling matrix $J^d$ contains at most
$\bar c^{2d}$ elements, each of them less than or equal to
$\|J\|^d$. Under the assumption of local finiteness,
we thus obtain the following Lieb-Robinson
bound~\cite{LiebCMP72,NachtergaeleNewTrends07,BravyiPRL06,OsbornePRL06}
\begin{eqnarray}
  \mathcal{R}(n)&\leq& |S| \left\|\hat A_S\right\|
  \sum_{d=n+1}^\infty
  \frac{\left(2t\mathcal{O}\bar c^2\left\|J\right\|e^\mu\right)^d}{d!}
  e^{-\mu d}\nonumber\\
  &\leq& |S| \left\|\hat A_S\right\|e^{-\mu(n-vt)}\,,\label{eq:LR_A}
\end{eqnarray}
where 
\[
v=2\left\|\mathcal O\right\|\bar c^2\left\|J\right\|e^\mu/\mu\,,
\]
and $|S|$ accounts for $\hat O_S$ acting on several system nodes.
The factor $e^{\mu n}$ with $\mu>0$ has been introduced in $v$ to
emphasize the exponential decay of the error with the number of
layers taken into account.
Minimizing $\mathcal{R}(n)$ as a function of $\mu$ leads to the
choice $\mu=1$ and hence Eq.~(6) of the main text. 

\section{quasi-finite resolution of quantum dynamics for continuous environments}

We start from the generic Hamiltonian, 
Eq.~(7) in the main text, describing the interaction of a
system with a continuous environment. 
The goal is to bound the error made by evolving a generic system
operator $\hat A_S(t)$ employing the surrogate Hamiltonian $\hat
H_{P_n}$, Eq.~(8) in the main text, instead of the full generator $\hat H$.
Using the unitary invariance of the norm, i.e., $\|\hat U \hat A \hat
U^+\|=\|\hat A\|$, and the triangle inequality, 
the error is expressed as
\begin{widetext}
\begin{eqnarray}
  \left\|\hat A_S(t)\!-\! \hat A_S^{H_{P_n}}(t)\right\|
  &=& \left\|e^{-i\hat H_{P_n}t}e^{i\hat Ht}
    \hat A_Se^{-i\hat Ht}e^{i\hat H_{P_n}}-\hat A_S\right\|\nonumber\\
  &\leq& \left\|e^{-i\hat H_{P_n}t}e^{i\hat Ht}
    \hat A_Se^{-i\hat Ht}e^{i\hat H_{P_n}}-e^{i(\hat H - \hat H_{P_n})t}\hat A_S
    e^{-i(\hat H-\hat H_{P_n})t}\right\|
  +\left\|e^{i(\hat H - \hat H_{P_n})t}\hat A_S
    e^{-i(\hat H-\hat H_{P_n})t}\!-\!\hat A_S\right\|\nonumber\\
  &\equiv&R_1(P_n)+R_2(P_n)\label{eq:aplusb}\,.
\end{eqnarray}  
\end{widetext}
The strategy is now to bound each of the two terms, $R_1(P_n)$ and
$R_2(P_n)$. 

We first consider $R_1(P_n)$ and define a function
$F(\lambda)$~\cite{DeRaedtCPR87},  
\begin{equation}
  F(\lambda)=1-e^{i\lambda \hat H}e^{-i\lambda \hat H_{P_n}}e^{-i\lambda(\hat H-\hat H_{P_n})},
\label{eq:F}\end{equation}
with $F(0)=0$. Derivation with respect to $\lambda$ yields
\begin{equation}
  \frac{\partial F(\lambda)}{\partial\lambda} = 
  e^{i\lambda \hat H} 
  \left[e^{-i\lambda \hat H_{P_n}}\;,\;\hat H\right]
  e^{-i\lambda(\hat  H-\hat H_{P_n})} \,.
  \label{eq:der}
\end{equation}
Applying the Kubo identity \cite{Kubo57}, 
\begin{displaymath}
  \left[\hat H \;,\;e^{-i\lambda \hat H_{P_n}}\right]
  = i\int_0^{\lambda}e^{-i(\lambda-\mu)\hat H_{P_n}}
  \left[\hat H\;,\;\hat H_{P_n}\right]e^{-i\mu\hat H_{P_n}}d\mu.
\end{displaymath}
we can rewrite Eq.~\eqref{eq:der}, obtaining 
\begin{widetext}
\begin{equation}
  \frac{\partial F(\lambda)}{\partial\lambda}
  =-i \int_0^\lambda\!\!d\mu
  e^{i\lambda \hat H}\!\!e^{-i\mu\hat H_{P_n}}\!\left[\hat H\;,\;\hat H_{P_n}\right]
  e^{-i(\lambda-\mu)\hat H_{P_n}}\!e^{-i\lambda(\hat H-\hat H_{P_n})}\,.
  \label{eq:der2}
\end{equation}
Integrating, using the initial condition $F(0)=0$ and
Eq.~\eqref{eq:F}, yields 
\begin{equation}
  e^{-i(\hat H-\hat H_{P_n})t}-e^{-i\hat Ht}e^{\hat H_{P_n}t}
  =-i\int_0^t\!d\lambda\!\!\int_0^\lambda d\mu
  e^{i\lambda \hat H}e^{-i\mu\hat H_{P_n}}\![\hat H,\hat H_{P_n}]
  e^{-i(\lambda-\mu)\hat H_{P_n}}e^{-i\lambda(\hat H-\hat H_{P_n})}\,.
\label{eq:sn}\end{equation}
Estimation of the norms and of the integrals in Eq.~\eqref{eq:sn} leads to~\cite{DeRaedtCPR87,PoulinPRL11} 
\begin{equation}
  \left\|e^{-i(\hat H-\hat H_{P_n})t} - 
    e^{-i\hat Ht}e^{i\hat H_{P_n}t}\right\| \leq
  \frac{t^2}{2}\left\|\left[\hat H\;,\;\hat H_{P_n}\right]\right\|\,.
\label{eq:deraedt}
\end{equation} 
Equation~\eqref{eq:deraedt} allows the following estimate for $R_1(P_n)$
\begin{eqnarray}
  R_1(P_n)&=&\left\|e^{-i\hat H_{P_n}t}e^{i\hat Ht}\hat
    A_Se^{-i\hat Ht}e^{i\hat H_{P_n}}-e^{i(\hat H - \hat H_{P_n})t}\hat A_S
    e^{-i(\hat H-\hat H_{P_n})t}\right\|
  \nonumber\\ &\leq&
  2\left\|e^{-i\hat H_{P_n}t}e^{i\hat Ht}\hat
    A_S\left(e^{-i\hat Ht}e^{i\hat H_{P_n}}-e^{-i(\hat H - \hat H_{P_n})t}\right)
    \right\|\nonumber\\
  &&+\left\|\left(e^{i\hat Ht}e^{-i\hat H_{P_n}}-e^{i(\hat H - 
  \hat H_{P_n})t}\right)\hat A_S
    \left(e^{-i\hat Ht}e^{i\hat H_{P_n}}-e^{-i(\hat H - 
    \hat H_{P_n})t}\right)\right\|\nonumber\\
  &\leq&\left\|\hat A_S\right\|
  \frac{t^2}{2}\left\|\left[\hat H\;,\;\hat H_{P_n}\right]\right\|
  \left(2+\frac{t^2}{2}\left\|\left[\hat H\;,\;\hat H_{P_n}\right]\right\|\right).
  \label{eq:A}
\end{eqnarray}
\end{widetext}
We thus find that $R_1(P_n)$ is bounded by a second order polynomial  in 
$t^2\left\|[\hat H,\hat H_{P_n}]\right\|$. For finite $t$, $R_1(P_n)$
is determined by 
the  commutator of $\hat H$ and $\hat H_{P_n}$ which 
vanishes in the limit $n\rightarrow\infty$. In particular, we find
\begin{widetext}
\begin{eqnarray}
  \left[\hat H\;,\;\hat H_{P_n}\right] &=&
  \left[\hat H_S\;,\;\hat O^I_S\right]
  \left\{\sum_{i=0}^{n-1}\tilde J_i
    \left(\hat c_i+\hat{c}^+_i\right)-
    \int_{0}^{x_{max}}J(x) \left(\hat c_x+\hat c_x^+\right)dx\right\}
  \nonumber \\
  &&+\left(\hat O^I_S\right)^2\left[ 
    \int_{0}^{x_{max}}J(x)(\hat c_x+\hat c_x^+)dx 
    \;,\; \sum_{i=0}^{n-1}\tilde J_i(\hat c_i+\hat{c}^+_i)\right] 
  +\hat O^I_S\left[\int_{0}^{x_{max}}J(x)\left(
      \hat c_x+\hat c_x^+\right)dx \;,\;
      \sum_{i=0}^{n-1}\tilde{g}_i 
      \hat c^+_i\hat{c}_i\right]\nonumber\\
    &&+\left[\int_{0}^{x_{max}}g(x)
      \hat c^+_x\hat c_x dx \;,\; \sum_{i=0}^{n-1}\tilde J_i
      \left(\hat c_i+\hat{c}^+_i\right)\right]\hat O^I_S
    +\left[\int_{0}^{x_{max}}g(x)
      \hat c^+_x\hat c_x dx \;,\;
      \sum_{i=0}^{n-1}\tilde{g}_i \hat c^+_i\hat{c}_i\right]\nonumber\\
    &&+\hat O^I_S\left[
      \int_{0}^{x_{max}}J(x)(\hat c_x+\hat c_x^+)dx \;,\; 
      2\sum_{i<j=0}^{n-1}\tilde{K}_{ij} \left(
        \hat c_i\hat c^+_j+\hat c_i\hat c^+_j
        +\hat c^+_i\hat c_i\hat c^+_j\hat c_j\right)\right]\nonumber\\
    &&+\left[
      2\int_0^{x_{max}}\int_x^{x_{max}} K\left(|x-x'|\right)
      \left(\hat c_x\hat c^+_{x'}+ \hat c_x^+\hat c_{x'}
        +\hat c^+_x\hat c_x\hat c^+_{x'}\hat c_{x'}\right) dxdx' \;,\;
        \sum_{i=0}^{n-1}\tilde J_i \left(\hat c_i+\hat{c}^+_i\right)
      \right]\hat O^I_S\nonumber\\
      &&+4\left[\int_0^{x_{max}}\int_x^{x_{max}}
        K\left(|x-x'|\right)\left(\hat c_x\hat c^+_{x'}
          +\hat c_x^+\hat c_{x'}
          +\hat c^+_x\hat c_x\hat c^+_{x'}\hat c_{x'}\right) 
        dxdx' \;,\; \sum_{i<j=0}^{n-1}\tilde{K}_{ij}\left(
          \hat c_i\hat c^+_j+\hat c_i^+\hat c_j
          +\hat c^+_i\hat c_i\hat c^+_j\hat c_j\right)\right]\nonumber\\
      &&+2\left[\int_0^{x_{max}}\int_x^{x_{max}}
        K\left(|x-x'|\right)
        \left(\hat c_x\hat c^+_{x'}+\hat c_x^+\hat c_{x'}
          +\hat c^+_x\hat c_x\hat c^+_{x'}\hat c_{x'}\right)
        dxdx' \;,\; \sum_{i=0}^{n-1}\tilde{g}_i 
        \hat c^+_i\hat{c}_i\right]\nonumber\\
      &&+\left[\int_{0}^{x_{max}}g(x)
        \hat c^+_x\hat c_x dx \;,\;
        2\sum_{i<j=0}^{n-1}\tilde{K}_{ij}\left(
          \hat c_i\hat c^+_j+\hat c^+_i\hat c_j 
          +\hat c^+_i\hat c_i\hat c^+_j\hat c_j\right)\right]\,.
    \label{eq:comm}
\end{eqnarray}
Any commutator acting on the bath degrees of freedom has the generic
form $[\hat A_x,\hat B_{x'}]=[\hat A_x,\hat B_{x}]\delta_{x,x'}$. 
Therefore commutators between local bath operators vanish, 
\begin{eqnarray*}
  \left(\hat O^I_S\right)^2
  \left[ \int_{0}^{x_{max}}J(x)
    \left(\hat c_x+\hat c_x^+\right)dx\;,\;
    \sum_{i=0}^{n-1}\tilde J_i
    (\hat c_i+\hat{c}^+_i)\right] &=&0 \,,\\  
  \left[\int_{0}^{x_{max}}g(x)
    \hat c^+_x \hat c_x dx \;,\; 
    \sum_{i=0}^{n-1}\tilde{g}_i \hat c^+_i\hat{c}_i\right]&=&0\,,\\
  \hat O^I_S\left[\int_{0}^{x_{max}}J(x)
    (\hat c_x+\hat c_x^+)dx \;,\;
    \sum_{i=0}^{n-1}\tilde{g}_i \hat c^+_i\hat{c}_i\right]
  +\left[\int_{0}^{x_{max}}g(x)
    \hat c^+_x \hat c_x dx \;,\;
    \sum_{i=0}^{n-1}\tilde J_i\left(\hat c_i+\hat{c}^+_i\right)
  \right]\hat O^I_S&=&0\,,
\end{eqnarray*}
and we are left with those terms in Eq.~\eqref{eq:comm} that involve
non-local bath operators,
\begin{eqnarray}
  \label{JK}\hat C_{JK} &=& 
  \hat O^I_S\left[\int_{0}^{x_{max}}J(x)
    \left(\hat c_x+\hat  c_x^+\right)dx \;,\;
    2\sum_{i<j=0}^{n-1}
    \tilde{K}_{ij}\left(\hat c_i\hat c^+_j+\hat c_i^+\hat c_j
      +\hat c^+_i\hat c_i\hat c^+_j\hat c_j\right)\right] \\
  &&-2\left[\int_0^{x_{max}}\int_x^{x_{max}}K\left(|x-x'|\right)
    \left(\hat c_x\hat c^+_{x'}+ \hat c_x^+\hat c_{x'}+
      \hat c^+_x\hat c_x\hat c^+_{x'}\hat c_{x'}\right) dxdx'\;,\;
    \sum_{i=0}^{n-1}\tilde J_i
    \left(\hat c_i+\hat{c}^+_i\right)\right]\hat O^I_S \,,\nonumber \\
  \label{KK} \hat C_{K} &=&\\
  &&  4\left[
    \int_0^{x_{max}}\int_x^{x_{max}}K\left(|x-x'|\right)
    \left(\hat c_x\hat c^+_{x'}+\hat c_x^+\hat c_{x'}
      +\hat c^+_x\hat c_x\hat c^+_{x'}\hat c_{x'}\right) dxdx' \;,\;
    \sum_{i<j=0}^{n-1}\tilde{K}_{ij}\left(
      \hat c_i\hat c^+_j+\hat c_i^+\hat c_j
      +\hat c^+_i\hat c_i\hat c^+_j\hat c_j\right)\right]\,,\nonumber\\
  \label{GK}\hat C_{gK} 
  &=& 2\left[\int_0^{x_{max}}\int_x^{x_{max}}K\left(|x-x'|\right)
    \left(\hat c_x\hat c^+_{x'}+\hat c_x^+\hat c_{x'}
      +\hat c^+_x\hat c_x\hat c^+_{x'}\hat c_{x'}\right) dxdx'\;,\;
    \sum_{i=0}^{n-1}\tilde{g}_i \hat c^+_i\hat{c}_i\right]\\
  &&-2\left[\int_{0}^{x_{max}}g(x)
    \hat c^+_x\hat c_x dx \;,\;
    \sum_{i<j=0}^{n-1}\tilde{K}_{ij}\left(
      \hat c_i\hat c^+_j+\hat c_i^+\hat c_j
      +\hat c^+_i\hat c_i\hat c^+_j\hat c_j\right)
  \right]\,.\nonumber
\end{eqnarray}
Using these definitions of $\hat C_{JK}$, $\hat C_K$, $\hat
C_{gK}$ and the triangular inequality, the norm of the commutator of
the full generator and the surrogate one can be rewritten,
\begin{eqnarray}
\left\|[\hat H,\hat H_{P_n}]\right\| \leq
\left\|[\hat H_S,\hat O^I_S]\right\|2R_J(P_n)
+\!\|\hat C_{JK}\|\!+\!\|\hat C_{KK}\|\!+\!\|\hat C_{gK}\| \,,
\label{eq:norm}
\end{eqnarray}
where 
\begin{displaymath}
R_J(P_n)\!\leq\!\sum_i\left(
\left\| J(x_i)\delta x_i(c_i+c^+ _i)\!-\!\!\int_{\delta x_i}J(x)(c_x+c^+_x)dx\right\|\right)
\end{displaymath}
bounds the error made by evaluating the integral over
$J(x)(c_x+c^+_x)$ in terms of the Riemann sum built on  
the partition $P_n$. This error vanishes in the limit
$|P_n|\rightarrow 0$, where $|P_n|=\max_{i\leq n}\delta x_i$.  The
norm of the remaining terms in Eq.~\eqref{eq:norm}  
can be evaluated analogously. To this end, we rewrite $\hat C_{JK}$ in
Eq.~\eqref{JK}, 
\begin{eqnarray}
  \hat C_{JK}&=&2\hat O^I_S\sum_{i\neq j}J(x_i)\tilde{K}_{ij}
  \left[\hat c_i+\hat c^+_i \;,\;
    \hat c_i\hat c^+_j+\hat c_i^+\hat c_j
    +\hat c^+_i\hat c_i\hat c^+_j\hat c_j\right]\\
  &&-2\hat O^I_S\int_0^{x_{max}} \sum_{i:x_i\neq x}
  K\left(|x-x_i|\right)
  \tilde{J_i}\left[\hat c_i+\hat c^+_i \;,\;
    \hat c_i\hat c^+_x+ \hat c_i^+\hat c_x+
    +\hat c^+_i\hat c_i\hat c^+_x\hat c_x\right] dx\,,\nonumber
\end{eqnarray}
such that we can estimate
\begin{equation}
\|C_{JK}\|\leq2\|\hat O^I_S\|R_{JK}(P_n)\,,\label{eq:RJK}
\end{equation}
where
\begin{eqnarray}
  R_{JK}(P_n)&=&\sum_{j=0}^{n-1}\sum_{i\neq j}
  \delta x_i\Bigg\|\int_{\delta x_j} 
  K\left(|x-x_i|\right)J(x_i)
  \left[\hat c_i+\hat c^+_i \;,\;
    \hat c_i\hat c^+_x+\hat c_i^+\hat c_x 
    +\hat c^+_i\hat c_i\hat c^+_x\hat c_x\right]dx \\
  &&\quad\quad\quad\quad-\delta x_j 
  K\left(|x_j-x_i|\right)J(x_i)
  \left[\hat c_i+\hat c^+_i \;,\;
    \hat c_i\hat c^+_j + \hat c_i^+\hat c_j +
    \hat c^+_i\hat c_i\hat c^+_j\hat c_j\right] \Bigg\|\nonumber
\end{eqnarray} 
bounding the error made by evaluating the integral over $x$ by 
the corresponding Riemann sum over $P_n$. 
Analogously we obtain
\begin{equation}
  \|\hat C_{gK}\|\leq 2R_{gK}(P_n) \label{eq:RgK}
\end{equation}
with 
\begin{eqnarray}
  R_{gK}(P_n)&=&\sum_{j=0}^{n-1}\sum_{i\neq j}
  \delta x_i\Bigg\|\int_{\delta x_j}g(x_i)K\left(|x-x_i|\right)
  \left[\hat c^+_i\hat c_{x'}+
    \hat c^+_i\hat c_{i}\hat c^+_{x'}\hat  c_{x'}\;,\;
    \hat c^+_i\hat c_{i}\right]dx \\
  &&\quad\quad\quad\quad
  -\delta x_j g(x_i)K(|x_i-x_j|)\left[\hat c^+_i\hat c_{j} \nonumber
    +\hat c^+_i\hat c_{i}\hat c^+_{j}\hat c_{j}\;,\;
    \hat c^+_i\hat c_{i}\right]\Bigg\|
\end{eqnarray}
and 
\begin{equation}
  \hat C_{K}=4\int_0^{x_{max}}dx
  \sum_{\substack{i\neq j\\i:x_i\neq x}}
  K(|x-x_i|)\tilde{K}_{ij}
  \left(\hat d_1+\hat d_2+\hat d_3+\hat d_4+\hat d_5\right)\,,
  \label{eq:ck}
\end{equation}
\end{widetext}
where 
\begin{subequations}\label{eq:diffs}
  \begin{eqnarray}
    \hat d_1&=&\hat c_x\left[\hat c^+_i\;,\;\hat c_i\right]\hat c^+_j
    -\hat c_j\left[\hat c^+_i\;,\;\hat c_i\right]\hat c^+_x \,,\\
    \hat d_2&=& \hat c^+_j\hat c_j
    \left[\hat c_i\;,\;\hat c^+_i\hat c_i\right]c^+_{x}
    -\hat c_x\hat c^+_{x}\left[\hat c_i\;,\;\hat c^+_i\hat
      c_i\right]c^+_{j}
    \,,\\
    \hat d_3&=& \left[\hat c_i\;,\;\hat c^+_i\hat c_i\right]
    \left(\hat c^+_j\hat c_jc^+_x-\hat c^+_x\hat c_xc^+_j\right)\,,\\
    \hat d_4&=&\hat c_x
    \left[\hat c^+_i\;,\;\hat c^+_i\hat c_i\right]\hat c^+_j\hat c_j
    -\hat c_j\left[\hat c^+_i\;,\;\hat c^+_i\hat c_i\right]\hat
    c^+_x\hat c_x\,,\\ 
    \hat d_5&=&\left(\hat c^+_j\hat c_jc^+_x
      -\hat c^+_x\hat  c_xc^+_j\right)
    \left[\hat c_i\;,\;\hat c^+_i\hat c_i\right]\,.
\end{eqnarray}  
\end{subequations}
Equations~\eqref{eq:diffs} imply that $\hat C_K$
depends on the partition along $x$, i.e., it would vanish if
$\sum_j\rightarrow \int dx$. 
Its norm can consequently be bounded as
\begin{equation}
  \|\hat C_{K}\|\leq 2R_{K}(P_n)
  \label{eq:RKK}
\end{equation}
where 
\begin{widetext}
  \begin{eqnarray}
    R_{K}(P_n)=5\sum_j\sum_{i\neq j}\left|
      \delta x_i\delta x_j\int_{\delta x_j}
      K\left(|x-x_i|\right)K\left(|x_i-x_j|\right)
      \max (\|\hat d_1\|,\|\hat d_2\|,\|\hat d_3\|,\|\hat d_4\|,
      \|\hat d_5\|)\right|dx.
\end{eqnarray}
Using Eqs.~\eqref{eq:norm}, \eqref{eq:RJK}, \eqref{eq:RgK} and
\eqref{eq:RKK}, the final estimate can be written
\begin{eqnarray}
  \left\|[\hat H,\hat H_{P_n}]\right\|&\leq&
  2\left(
    \left\|\left[\hat H_S\;,\;\hat O^I_S\right]\right\|R_J(P_n)
    +\left\|\hat A_S\right\|R_{JK}(P_n)+R_{gK}(P_n)+R_{K}(P_n)
  \right)\nonumber\\
  &=&2\left(
    \left\|\left[\hat H_S\;,\;\hat O^I_S\right]\right\|R_J(P_n)
    +R_B(P_n)\right)\,,\label{eq:R12}
\end{eqnarray}
\end{widetext}
where $R_B(P_n)$ comprises of all the errors due to discretization
of the integrals involving $K(|x-x'|)$, $J(x)K(|x-x'|)$ and
$g(x)K(|x-x'|)$. 
Since $J(x)$ represents energy relaxation, $R_J(P_n)$
vanishes for pure dephasing. Similarly, $R_B(P_n)$ captures 
the intra-bath interactions and vanishes for normal modes.  Using
Eqs~\eqref{eq:R12} and \eqref{eq:A}  one obtains
\begin{widetext}
\begin{equation}
R_1(P_n)\leq\left\|\hat A_S\right\|
  t^2\left(
    \left\|\left[\hat H_S\;,\;\hat O^I_S\right]\right\|R_J(P_n)
    +R_B(P_n)\right)
  \left[2+t^2\left(
    \left\|\left[\hat H_S\;,\;\hat O^I_S\right]\right\|R_J(P_n)
    +R_B(P_n)\right)\right]
\label{eq:R1}\end{equation}
\end{widetext} 
thus showing that at finite times the error $R_1(P_n)$ 
depends on how well the integrals are approximated
by the Riemann sums.

The second contribution to the error, $R_2(P_n)$ in
Eq.~\eqref{eq:aplusb}, represents the distance between $\hat A_S$ and
its time evolution under $\hat H-\hat H_{P_n}$. It can be estimated 
\begin{eqnarray}
  R_2(P_n)\!&=&\!\!\left\|e^{i(\hat H - \hat H_{P_n})t}\hat A_S
    e^{-i(\hat H-\hat H_{P_n})t}-\hat A_S\right\|\!\!\nonumber\\
  &&\leq\sum_{k=1}^{\infty}\frac{t^k}{k!}\!\left\|\hat
    C_k^{H-H_{P_n}}\right\|
  \label{eq:BCHR2}
\end{eqnarray}
with 
\[
\hat C^{H-H_{P_n}}_k =
\left[\hat H-\hat H_{P_n} \;,\; \hat C^{H-H_{P_n}}_{k-1}\right]
\]
and $\hat C^{H-H_{P_n}}_0=\hat O_S$
using the Baker-Campbell-Hausdorff formula and
the triangular inequality.
Since
\begin{displaymath}
  \left\|\hat C_1^{\hat H-\hat H_{P_n}}\right\|
  =\left\|\left[\hat H-\hat H_{P_n} \;,\;\hat A_S\right]\right\|
  \leq2\left\|\hat A_S\right\|\left\|\hat H-\hat H_{P_n} \right\|\,,
\end{displaymath}
we obtain for $\hat C^{\hat H-\hat H_{P_n}}_k$, 
\begin{equation}
  \left\|\hat C^{H-H_{P_n}}_k\right\| \leq 
 2^k \left\|\hat A_S\right\|\left\|\hat H-\hat H_{P_n}\right\|^k.
\end{equation}
Substituting this into Eq.~\eqref{eq:BCHR2} yields the following
estimate for $R_2(P_n)$, 
\begin{equation}
  R_2(P_n)\leq\left\|\hat A_S\right\|
  \left(e^{2\|\hat H-\hat H_{P_n}\|t}-1\right)\label{eq:R2}.
\end{equation}
For finite time, the error $R_2(P_n)$ vanishes in the limit
$n\rightarrow\infty$.

We conclude from Eqs.~\eqref{eq:A}, \eqref{eq:R12} and \eqref{eq:R2}
that, for a fixed finite time $t$, the error $R(P_n)$ can be made
arbitrarily small by proper choice of the partition. It is thus
sufficient to represent a continuous bath with infinitely many DOF
by a finite set of 'surrogate' modes. Note that in our derivation no
assumption on the system-bath interaction or intra-bath couplings were
made. 

As a final remark we note that, as long as the full Hamiltonian
contains bounded operators, the bounds, 
Eq.~\eqref{eq:LR_A} for discrete DOF and Eq.~\eqref{eq:R2} for continuous
DOF, depend only on the coupling structure and not the specific
algebraic form of $\hat H$. 

\section{The Suzuki-Trotter Decomposition}
At finite $t$ the effective generator  of the reduced system evolution
has the  generic form 
\begin{displaymath} 
\hat H_{X_n} = \sum_{i,j\in X_n}\hat h_{ij}.
\end{displaymath}
where without loss of generality we assume two-body interactions. The
set $X_n$ is that of the relevant DOF on which $\hat H_{X_n}$ acts
and $\hat h_{ij}$ is the generic interaction between two DOF. At
$t<\infty$  the effective propagator generated by  $\hat H_{X_n}$
can be approximated by applying a Suzuki-Trotter
expansion~\cite{Trotter59,SuzukiCMP94}  
\begin{equation}
  e^{-i\hat H_{X_n}t}\approx   
  \left( \prod_{\{i,j\}\in X_n}e^{-i\hat h_{ij}\Delta t}\right)^{m_n}\,,
  \label{eq:trott}
\end{equation}
where $\Delta t = t/m_n$. 
The generator $\hat H_{X_n}$ contains $K_n\leq|X_n|^2$ two-body terms.
The error introduced by approximating $e^{-i\hat H_n\Delta t}$
within each $\Delta t$  by a product of $K_n$ terms
is of the order 
$\epsilon_2\leq\frac{1}{2}\mathcal{O}^2K_{n}^2(\Delta t)^2$ 
\cite{DeRaedtCPR87,PoulinPRL11}. A prespecified error $\epsilon_2/2$ for
the whole time $t$ is 
achieved by taking $m_n=\mathcal{O}^2 t^2 K_{n}^2/\epsilon_2$ Trotter
steps, i.e., by choosing $\Delta t=\epsilon_2/(t\mathcal{O}^2 K_{n}^2)$.
The product formula in Eq.~\eqref{eq:trott} can be generalized to 
generators exhibiting arbitrary 
time-dependence~\cite{PoulinPRL11}. 

\section{Extension to $k$-body interactions}
\subsection{Extension of the dynamical bounds}
A generic generator defined on a discrete set and containing $k$-body
interactions is written as 
\begin{equation}
  \hat H=\sum_{i_1,i_2,\cdots,i_k}\hat h_{i_1,i_2,\cdots,i_k}\,,\label{eq:kbo}
\end{equation}
where  $i_1,\cdots,i_k\in(0,\infty)$, and $\hat h_{i_1,\cdots,i_k}$ is
a generic $k$-body interaction. 
This Hamiltonian defines a hypergraph,  i.e., an ordered pair $G=(N,E)$
with the set  of nodes $N$ made up of all the degrees of freedom
$\hat H$ acts upon  and $E$ comprising the set of 
non-empty subsets of $N$, called hyperedges or links, for which
$\|\hat h_{i_1,\cdots,i_k}\|\neq0$. Since all  interactions are
$k$-local in Eq.~\eqref{eq:kbo}, all hyperedges have size 
$k$, and the hypergraph is $k$-uniform. A graph can therefore be
regarded as a 2-uniform hypergraph. The adjacency matrix $A^h_{ij}$ of
a hypergraph $G$ is defined as the matrix whose entries $A^h_{ij}$
correspond to the number of hyperedges containing both degrees of
freedom $i$ and $j$~\cite{RodriguezAML09}.  
The connectivity of a node $c_i$ is  given by the number of hyperedges
involving the node, $c_i=\sum_jA^{h}_{ij}$. The hypergraph is therefore
{\it locally finite} if $\max_{i\in N}c_i=\bar c_i<\infty$. One can
then define the coupling matrix $J$ on the hypergraph, 
\begin{displaymath}
  J_{ij}=\sqrt{\sum_{\mu\nu}\left(
      \sum_{\substack{i_1,\cdots,i_k : \\ \exists (k,k') : i_k=i,
          i_{k'}=j}}
      [J^{\mu\nu}_{i_1,\cdots,i_k}]^2\right)}\,,
\end{displaymath}
and consequently bound its  norm by $\|J\|\leq\max_{i\in N}\sum
J_{ij}$. This implies that Eq. (6) in the main text holds in the same
form with $\mathcal O=\max_{i_1,\cdots,i_k}\|\hat h_{i_1,\cdots,i_k}\|$.
Equation (9) in the main text holds formally unaltered as well, with
the Riemann sums calculated for the relative $k$-body terms.

\subsection{Suzuki-Trotter decomposition}
The propagator for a $k$-body effective generator of the form
Eq.~\eqref{eq:kbo} is decomposed as  
\begin{displaymath}
  e^{-i\hat H_{X_n}t}\approx \left(
    \prod_{\{i_1,\cdots,i_k\}\in X_n}e^{-i\hat h_{i_1,\cdots,i_k}t}\right)^{m_n}\,.
\end{displaymath}
The error estimate in the previous section holds formally unaltered
with  $K_n\leq|X_n|^k$. One could then use the
Solovay-Kitaev algorithm~\cite{DawsonQIC06} to further
decompose each $k$-unitary into a product of one- and two-body
unitaries chosen from a suitable set. To achieve an accuracy $\e$
for each $k$-unitary transformation, $n_{SK}=a\log_2^b(\e^{-1})$
operations are required with $a$, $b$ constants. If one chooses
$\e=\e_2/(2n_d)$,  the effective propagator is simulated with an
accuracy $\e_2$ employing $n'_d=a n_d\log_2^b(n_d/\e_2)$ one- and two-body
unitaries, i.e., with a computational effort that scales polynomially
in time and number of effective DOF~\cite{PoulinPRL11}.



\begin{thebibliography}{40}
\expandafter\ifx\csname natexlab\endcsname\relax\def\natexlab#1{#1}\fi
\expandafter\ifx\csname bibnamefont\endcsname\relax
  \def\bibnamefont#1{#1}\fi
\expandafter\ifx\csname bibfnamefont\endcsname\relax
  \def\bibfnamefont#1{#1}\fi
\expandafter\ifx\csname citenamefont\endcsname\relax
  \def\citenamefont#1{#1}\fi
\expandafter\ifx\csname url\endcsname\relax
  \def\url#1{\texttt{#1}}\fi
\expandafter\ifx\csname urlprefix\endcsname\relax\def\urlprefix{URL }\fi
\providecommand{\bibinfo}[2]{#2}
\providecommand{\eprint}[2][]{\url{#2}}

\bibitem[{\citenamefont{H.-P-Breuer and Petruccione}(2007)}]{Breuer}
\bibinfo{author}{\bibnamefont{H.-P-Breuer}} \bibnamefont{and}
  \bibinfo{author}{\bibfnamefont{F.}~\bibnamefont{Petruccione}},
  \emph{\bibinfo{title}{The Theory of Open Quantum Systems}}
  (\bibinfo{publisher}{Oxford Univ. Press}, \bibinfo{year}{2007}).

\bibitem[{\citenamefont{Barreiro et~al.}(2011)\citenamefont{Barreiro, M\"uller,
  Schindler, Nigg, Monz, Chwalla, Hennrich, Roos, Zoller, and
  Blatt}}]{BarreiroNat11}
\bibinfo{author}{\bibfnamefont{J.~T.} \bibnamefont{Barreiro}},
  \bibinfo{author}{\bibfnamefont{M.}~\bibnamefont{M\"uller}},
  \bibinfo{author}{\bibfnamefont{P.}~\bibnamefont{Schindler}},
  \bibinfo{author}{\bibfnamefont{D.}~\bibnamefont{Nigg}},
  \bibinfo{author}{\bibfnamefont{T.}~\bibnamefont{Monz}},
  \bibinfo{author}{\bibfnamefont{M.}~\bibnamefont{Chwalla}},
  \bibinfo{author}{\bibfnamefont{M.}~\bibnamefont{Hennrich}},
  \bibinfo{author}{\bibfnamefont{C.~F.} \bibnamefont{Roos}},
  \bibinfo{author}{\bibfnamefont{P.}~\bibnamefont{Zoller}}, \bibnamefont{and}
  \bibinfo{author}{\bibfnamefont{R.}~\bibnamefont{Blatt}},
  \bibinfo{journal}{Nature} \textbf{\bibinfo{volume}{470}},
  \bibinfo{pages}{486} (\bibinfo{year}{2011}).

\bibitem[{\citenamefont{Genes et~al.}(2009)\citenamefont{Genes, Mari, Vitali,
  and Tombesi}}]{GenesAdvAMOP09}
\bibinfo{author}{\bibfnamefont{C.}~\bibnamefont{Genes}},
  \bibinfo{author}{\bibfnamefont{A.}~\bibnamefont{Mari}},
  \bibinfo{author}{\bibfnamefont{D.}~\bibnamefont{Vitali}}, \bibnamefont{and}
  \bibinfo{author}{\bibfnamefont{P.}~\bibnamefont{Tombesi}},
  \bibinfo{journal}{Adv. in At., Mol. and Opt. Phys.}
  \textbf{\bibinfo{volume}{57}}, \bibinfo{pages}{33} (\bibinfo{year}{2009}).

\bibitem[{\citenamefont{Eichler et~al.}(2011)\citenamefont{Eichler, Moser,
  Chaste, Zdrojek, Wilson-Rae, and Bachtold}}]{EichlerNatNano11}
\bibinfo{author}{\bibfnamefont{A.}~\bibnamefont{Eichler}},
  \bibinfo{author}{\bibfnamefont{J.}~\bibnamefont{Moser}},
  \bibinfo{author}{\bibfnamefont{J.}~\bibnamefont{Chaste}},
  \bibinfo{author}{\bibfnamefont{M.}~\bibnamefont{Zdrojek}},
  \bibinfo{author}{\bibfnamefont{I.}~\bibnamefont{Wilson-Rae}},
  \bibnamefont{and} \bibinfo{author}{\bibfnamefont{A.}~\bibnamefont{Bachtold}},
  \bibinfo{journal}{Nature Nanotech.} \textbf{\bibinfo{volume}{6}},
  \bibinfo{pages}{339} (\bibinfo{year}{2011}).

\bibitem[{\citenamefont{Sarovar et~al.}(2010)\citenamefont{Sarovar, Ishizaki,
  Fleming, and Whaley}}]{SarovarNatPhys10}
\bibinfo{author}{\bibfnamefont{M.}~\bibnamefont{Sarovar}},
  \bibinfo{author}{\bibfnamefont{A.}~\bibnamefont{Ishizaki}},
  \bibinfo{author}{\bibfnamefont{G.~R.} \bibnamefont{Fleming}},
  \bibnamefont{and} \bibinfo{author}{\bibfnamefont{K.~B.}
  \bibnamefont{Whaley}}, \bibinfo{journal}{Nature Phys.}
  \textbf{\bibinfo{volume}{6}}, \bibinfo{pages}{462} (\bibinfo{year}{2010}).

\bibitem[{\citenamefont{de~Lange et~al.}(2010)\citenamefont{de~Lange, Wang,
  Rist\'e, Dobrovitski, and Hanson}}]{deLangeSci10}
\bibinfo{author}{\bibfnamefont{G.}~\bibnamefont{de~Lange}},
  \bibinfo{author}{\bibfnamefont{Z.~H.} \bibnamefont{Wang}},
  \bibinfo{author}{\bibfnamefont{D.}~\bibnamefont{Rist\'e}},
  \bibinfo{author}{\bibfnamefont{V.~V.} \bibnamefont{Dobrovitski}},
  \bibnamefont{and} \bibinfo{author}{\bibfnamefont{R.}~\bibnamefont{Hanson}},
  \bibinfo{journal}{Science} \textbf{\bibinfo{volume}{330}},
  \bibinfo{pages}{60} (\bibinfo{year}{2010}).

\bibitem[{\citenamefont{Bluhm et~al.}(2010)\citenamefont{Bluhm, Foletti, Neder,
  Rudner, Mahalu, Umansky, and Yacoby}}]{BluhmNatPhys10}
\bibinfo{author}{\bibfnamefont{H.}~\bibnamefont{Bluhm}},
  \bibinfo{author}{\bibfnamefont{S.}~\bibnamefont{Foletti}},
  \bibinfo{author}{\bibfnamefont{I.}~\bibnamefont{Neder}},
  \bibinfo{author}{\bibfnamefont{M.}~\bibnamefont{Rudner}},
  \bibinfo{author}{\bibfnamefont{D.}~\bibnamefont{Mahalu}},
  \bibinfo{author}{\bibfnamefont{V.}~\bibnamefont{Umansky}}, \bibnamefont{and}
  \bibinfo{author}{\bibfnamefont{A.}~\bibnamefont{Yacoby}},
  \bibinfo{journal}{Nature Phys.} \textbf{\bibinfo{volume}{7}},
  \bibinfo{pages}{109} (\bibinfo{year}{2010}).

\bibitem[{\citenamefont{Kolkowitz et~al.}(2012)\citenamefont{Kolkowitz,
  Unterreithmeier, Bennett, and Lukin}}]{KolkowitzPRL12}
\bibinfo{author}{\bibfnamefont{S.}~\bibnamefont{Kolkowitz}},
  \bibinfo{author}{\bibfnamefont{Q.~P.} \bibnamefont{Unterreithmeier}},
  \bibinfo{author}{\bibfnamefont{S.~D.} \bibnamefont{Bennett}},
  \bibnamefont{and} \bibinfo{author}{\bibfnamefont{M.~D.} \bibnamefont{Lukin}},
  \bibinfo{journal}{Phys. Rev. Lett.} \textbf{\bibinfo{volume}{109}},
  \bibinfo{pages}{137601} (\bibinfo{year}{2012}).

\bibitem[{\citenamefont{Taminiau et~al.}(2012)\citenamefont{Taminiau, Wagenaar,
  van~der Sar, Jelezko, Dobrovitski, and Hanson}}]{TaminiauPRL12}
\bibinfo{author}{\bibfnamefont{T.~H.} \bibnamefont{Taminiau}},
  \bibinfo{author}{\bibfnamefont{J.~J.~T.} \bibnamefont{Wagenaar}},
  \bibinfo{author}{\bibfnamefont{T.}~\bibnamefont{van~der Sar}},
  \bibinfo{author}{\bibfnamefont{F.}~\bibnamefont{Jelezko}},
  \bibinfo{author}{\bibfnamefont{V.~V.} \bibnamefont{Dobrovitski}},
  \bibnamefont{and} \bibinfo{author}{\bibfnamefont{R.}~\bibnamefont{Hanson}},
  \bibinfo{journal}{Phys. Rev. Lett.} \textbf{\bibinfo{volume}{109}},
  \bibinfo{pages}{137602} (\bibinfo{year}{2012}).

\bibitem[{\citenamefont{Breuer et~al.}(2009)\citenamefont{Breuer, Laine, and
  Piilo}}]{BreuerPRL09}
\bibinfo{author}{\bibfnamefont{H.-P.} \bibnamefont{Breuer}},
  \bibinfo{author}{\bibfnamefont{E.-M.} \bibnamefont{Laine}}, \bibnamefont{and}
  \bibinfo{author}{\bibfnamefont{J.}~\bibnamefont{Piilo}},
  \bibinfo{journal}{Phys. Rev. Lett.} \textbf{\bibinfo{volume}{103}},
  \bibinfo{pages}{210401} (\bibinfo{year}{2009}).

\bibitem[{\citenamefont{Piilo et~al.}(2008)\citenamefont{Piilo, Maniscalco,
  H\"ark\"onen, and Suominen}}]{PiiloPRL08}
\bibinfo{author}{\bibfnamefont{J.}~\bibnamefont{Piilo}},
  \bibinfo{author}{\bibfnamefont{S.}~\bibnamefont{Maniscalco}},
  \bibinfo{author}{\bibfnamefont{K.}~\bibnamefont{H\"ark\"onen}},
  \bibnamefont{and} \bibinfo{author}{\bibfnamefont{K.-A.}
  \bibnamefont{Suominen}}, \bibinfo{journal}{Phys. Rev. Lett.}
  \textbf{\bibinfo{volume}{100}}, \bibinfo{pages}{180402}
  (\bibinfo{year}{2008}).

\bibitem[{\citenamefont{Breuer and Vacchini}(2008)}]{BreuerPRL08}
\bibinfo{author}{\bibfnamefont{H.-P.} \bibnamefont{Breuer}} \bibnamefont{and}
  \bibinfo{author}{\bibfnamefont{B.}~\bibnamefont{Vacchini}},
  \bibinfo{journal}{Phys. Rev. Lett.} \textbf{\bibinfo{volume}{101}},
  \bibinfo{pages}{140402} (\bibinfo{year}{2008}).

\bibitem[{\citenamefont{Koch et~al.}(2008)\citenamefont{Koch, Gro\ss{}mann,
  Stockburger, and Ankerhold}}]{KochPRL08}
\bibinfo{author}{\bibfnamefont{W.}~\bibnamefont{Koch}},
  \bibinfo{author}{\bibfnamefont{F.}~\bibnamefont{Gro\ss{}mann}},
  \bibinfo{author}{\bibfnamefont{J.~T.} \bibnamefont{Stockburger}},
  \bibnamefont{and}
  \bibinfo{author}{\bibfnamefont{J.}~\bibnamefont{Ankerhold}},
  \bibinfo{journal}{Phys. Rev. Lett.} \textbf{\bibinfo{volume}{100}},
  \bibinfo{pages}{230402} (\bibinfo{year}{2008}).

\bibitem[{\citenamefont{Baer and Kosloff}(1997)}]{BaerJCP97}
\bibinfo{author}{\bibfnamefont{R.}~\bibnamefont{Baer}} \bibnamefont{and}
  \bibinfo{author}{\bibfnamefont{R.}~\bibnamefont{Kosloff}},
  \bibinfo{journal}{J. Chem. Phys.} \textbf{\bibinfo{volume}{106}},
  \bibinfo{pages}{8862} (\bibinfo{year}{1997}).

\bibitem[{\citenamefont{Koch et~al.}(2002)\citenamefont{Koch, Kl\"uner, and
  Kosloff}}]{KochJCP02}
\bibinfo{author}{\bibfnamefont{C.~P.} \bibnamefont{Koch}},
  \bibinfo{author}{\bibfnamefont{T.}~\bibnamefont{Kl\"uner}}, \bibnamefont{and}
  \bibinfo{author}{\bibfnamefont{R.}~\bibnamefont{Kosloff}},
  \bibinfo{journal}{J. Chem. Phys.} \textbf{\bibinfo{volume}{116}},
  \bibinfo{pages}{7983} (\bibinfo{year}{2002}).

\bibitem[{\citenamefont{Koch et~al.}(2003)\citenamefont{Koch, Kl\"uner, Freund,
  and Kosloff}}]{KochPRL03}
\bibinfo{author}{\bibfnamefont{C.~P.} \bibnamefont{Koch}},
  \bibinfo{author}{\bibfnamefont{T.}~\bibnamefont{Kl\"uner}},
  \bibinfo{author}{\bibfnamefont{H.-J.} \bibnamefont{Freund}},
  \bibnamefont{and} \bibinfo{author}{\bibfnamefont{R.}~\bibnamefont{Kosloff}},
  \bibinfo{journal}{Phys. Rev. Lett.} \textbf{\bibinfo{volume}{90}},
  \bibinfo{pages}{117601} (\bibinfo{year}{2003}).

\bibitem[{\citenamefont{Hogben et~al.}(2010)\citenamefont{Hogben, Hore, and
  Kuprov}}]{HogbenJCP10}
\bibinfo{author}{\bibfnamefont{H.~J.} \bibnamefont{Hogben}},
  \bibinfo{author}{\bibfnamefont{P.}~\bibnamefont{Hore}}, \bibnamefont{and}
  \bibinfo{author}{\bibfnamefont{I.}~\bibnamefont{Kuprov}},
  \bibinfo{journal}{J. Chem. Phys.} \textbf{\bibinfo{volume}{132}},
  \bibinfo{pages}{174101} (\bibinfo{year}{2010}).

\bibitem[{\citenamefont{Hughes et~al.}(2009)\citenamefont{Hughes, Christ, and
  Burghardt}}]{HughesJCP09}
\bibinfo{author}{\bibfnamefont{K.~H.} \bibnamefont{Hughes}},
  \bibinfo{author}{\bibfnamefont{C.~D.} \bibnamefont{Christ}},
  \bibnamefont{and}
  \bibinfo{author}{\bibfnamefont{I.}~\bibnamefont{Burghardt}},
  \bibinfo{journal}{J. Chem. Phys.} \textbf{\bibinfo{volume}{131}},
  \bibinfo{pages}{024109} (\bibinfo{year}{2009}).

\bibitem[{\citenamefont{Lieb and Robinson}(1972)}]{LiebCMP72}
\bibinfo{author}{\bibfnamefont{E.~H.} \bibnamefont{Lieb}} \bibnamefont{and}
  \bibinfo{author}{\bibfnamefont{D.~W.} \bibnamefont{Robinson}},
  \bibinfo{journal}{Commun. Math. Phys.} \textbf{\bibinfo{volume}{28}},
  \bibinfo{pages}{251} (\bibinfo{year}{1972}).

\bibitem[{\citenamefont{Nachtergaele and Sims}(2009)}]{NachtergaeleNewTrends07}
\bibinfo{author}{\bibfnamefont{B.}~\bibnamefont{Nachtergaele}}
  \bibnamefont{and} \bibinfo{author}{\bibfnamefont{R.}~\bibnamefont{Sims}}, in
  \emph{\bibinfo{booktitle}{New Trends in Mathematical Physics. Selected
  contributions of the XVth International Congress on Mathematical Physics}},
  edited by \bibinfo{editor}{\bibfnamefont{V.}~\bibnamefont{Sidoravicius}}
  (\bibinfo{publisher}{Springer}, \bibinfo{year}{2009}), pp.
  \bibinfo{pages}{591--614}.

\bibitem[{SM()}]{SM}
\bibinfo{note}{{Supplementary Material}}.

\bibitem[{\citenamefont{Mohar and Woess}(1989)}]{MoharBLMS89}
\bibinfo{author}{\bibfnamefont{B.}~\bibnamefont{Mohar}} \bibnamefont{and}
  \bibinfo{author}{\bibfnamefont{W.}~\bibnamefont{Woess}},
  \bibinfo{journal}{Bull. London Math. Soc.} \textbf{\bibinfo{volume}{21}},
  \bibinfo{pages}{209} (\bibinfo{year}{1989}).

\bibitem[{\citenamefont{Bravyi et~al.}(2006)\citenamefont{Bravyi, Hastings, and
  Verstraete}}]{BravyiPRL06}
\bibinfo{author}{\bibfnamefont{S.}~\bibnamefont{Bravyi}},
  \bibinfo{author}{\bibfnamefont{M.~B.} \bibnamefont{Hastings}},
  \bibnamefont{and}
  \bibinfo{author}{\bibfnamefont{F.}~\bibnamefont{Verstraete}},
  \bibinfo{journal}{Phys. Rev. Lett.} \textbf{\bibinfo{volume}{97}},
  \bibinfo{pages}{050401} (\bibinfo{year}{2006}).

\bibitem[{\citenamefont{Osborne}(2006)}]{OsbornePRL06}
\bibinfo{author}{\bibfnamefont{T.~J.} \bibnamefont{Osborne}},
  \bibinfo{journal}{Phys. Rev. Lett.} \textbf{\bibinfo{volume}{97}},
  \bibinfo{pages}{157202} (\bibinfo{year}{2006}).

\bibitem[{\citenamefont{Prior et~al.}(2010)\citenamefont{Prior, Chin, Huelga,
  and Plenio}}]{PriorPRL10}
\bibinfo{author}{\bibfnamefont{J.}~\bibnamefont{Prior}},
  \bibinfo{author}{\bibfnamefont{A.~W.} \bibnamefont{Chin}},
  \bibinfo{author}{\bibfnamefont{S.~F.} \bibnamefont{Huelga}},
  \bibnamefont{and} \bibinfo{author}{\bibfnamefont{M.~B.}
  \bibnamefont{Plenio}}, \bibinfo{journal}{Phys. Rev. Lett.}
  \textbf{\bibinfo{volume}{105}}, \bibinfo{pages}{050404}
  (\bibinfo{year}{2010}).

\bibitem[{\citenamefont{Chin et~al.}(2010)\citenamefont{Chin, \'{A}ngel Rivas,
  Huelga, and Plenio}}]{ChinJMP10}
\bibinfo{author}{\bibfnamefont{A.~W.} \bibnamefont{Chin}},
  \bibinfo{author}{\bibnamefont{\'{A}ngel Rivas}},
  \bibinfo{author}{\bibfnamefont{S.~F.} \bibnamefont{Huelga}},
  \bibnamefont{and} \bibinfo{author}{\bibfnamefont{M.~B.}
  \bibnamefont{Plenio}}, \bibinfo{journal}{J. Math. Phys.}
  \textbf{\bibinfo{volume}{51}}, \bibinfo{pages}{092109}
  (\bibinfo{year}{2010}).

\bibitem[{\citenamefont{Burrell and Osborne}(2007)}]{BurrellPRL07}
\bibinfo{author}{\bibfnamefont{C.~K.} \bibnamefont{Burrell}} \bibnamefont{and}
  \bibinfo{author}{\bibfnamefont{T.~J.} \bibnamefont{Osborne}},
  \bibinfo{journal}{Phys. Rev. Lett.} \textbf{\bibinfo{volume}{99}},
  \bibinfo{pages}{167201} (\bibinfo{year}{2007}).

\bibitem[{\citenamefont{Hastings}(2008)}]{HastingsPRB08}
\bibinfo{author}{\bibfnamefont{M.}~\bibnamefont{Hastings}},
  \bibinfo{journal}{Phys. Rev. B} \textbf{\bibinfo{volume}{77}},
  \bibinfo{pages}{144302} (\bibinfo{year}{2008}).

\bibitem[{\citenamefont{Eisert et~al.}(2010)\citenamefont{Eisert, Cramer, and
  Plenio}}]{EisertRMP10}
\bibinfo{author}{\bibfnamefont{J.}~\bibnamefont{Eisert}},
  \bibinfo{author}{\bibfnamefont{M.}~\bibnamefont{Cramer}}, \bibnamefont{and}
  \bibinfo{author}{\bibfnamefont{M.}~\bibnamefont{Plenio}},
  \bibinfo{journal}{Rev. Mod. Phys.} \textbf{\bibinfo{volume}{82}}
  (\bibinfo{year}{2010}).

\bibitem[{\citenamefont{Cramer et~al.}(2008)\citenamefont{Cramer, Serafini, and
  Eisert}}]{Cramer08}
\bibinfo{author}{\bibfnamefont{M.}~\bibnamefont{Cramer}},
  \bibinfo{author}{\bibfnamefont{A.}~\bibnamefont{Serafini}}, \bibnamefont{and}
  \bibinfo{author}{\bibfnamefont{J.}~\bibnamefont{Eisert}}, in
  \emph{\bibinfo{booktitle}{Quantum information and many body quantum
  systems}}, edited by
  \bibinfo{editor}{\bibfnamefont{M.}~\bibnamefont{Ericsson}} \bibnamefont{and}
  \bibinfo{editor}{\bibfnamefont{S.}~\bibnamefont{Montangero}}
  (\bibinfo{publisher}{Pisa: Edizioni della Normale}, \bibinfo{year}{2008}),
  pp. \bibinfo{pages}{55--72}.

\bibitem[{\citenamefont{Nachtergaele et~al.}(2010)\citenamefont{Nachtergaele,
  Schlein, Sims, Starr, and Zagrebnov}}]{NachtergaeleRMathP10}
\bibinfo{author}{\bibfnamefont{B.}~\bibnamefont{Nachtergaele}},
  \bibinfo{author}{\bibfnamefont{B.}~\bibnamefont{Schlein}},
  \bibinfo{author}{\bibfnamefont{R.}~\bibnamefont{Sims}},
  \bibinfo{author}{\bibfnamefont{S.}~\bibnamefont{Starr}}, \bibnamefont{and}
  \bibinfo{author}{\bibfnamefont{V.}~\bibnamefont{Zagrebnov}},
  \bibinfo{journal}{Rev. Math. Phys.} \textbf{\bibinfo{volume}{22}},
  \bibinfo{pages}{207} (\bibinfo{year}{2010}).

\bibitem[{\citenamefont{Trotter}(1959)}]{Trotter59}
\bibinfo{author}{\bibfnamefont{H.~F.} \bibnamefont{Trotter}},
  \bibinfo{journal}{Proc. Am. Math. Soc.} \textbf{\bibinfo{volume}{10}},
  \bibinfo{pages}{545} (\bibinfo{year}{1959}).

\bibitem[{\citenamefont{Suzuki}(1993)}]{SuzukiProc93}
\bibinfo{author}{\bibfnamefont{M.}~\bibnamefont{Suzuki}},
  \bibinfo{journal}{Proc. Japan. Acad.} \textbf{\bibinfo{volume}{69}},
  \bibinfo{pages}{161} (\bibinfo{year}{1993}).

\bibitem[{\citenamefont{De~Raedt}(1987)}]{DeRaedtCPR87}
\bibinfo{author}{\bibfnamefont{H.}~\bibnamefont{De~Raedt}},
  \bibinfo{journal}{Comp. Phys. Rep.} \textbf{\bibinfo{volume}{7}},
  \bibinfo{pages}{1} (\bibinfo{year}{1987}).

\bibitem[{\citenamefont{Poulin et~al.}(2011)\citenamefont{Poulin, Qarry, Somma,
  and Verstraete}}]{PoulinPRL11}
\bibinfo{author}{\bibfnamefont{D.}~\bibnamefont{Poulin}},
  \bibinfo{author}{\bibfnamefont{A.}~\bibnamefont{Qarry}},
  \bibinfo{author}{\bibfnamefont{R.}~\bibnamefont{Somma}}, \bibnamefont{and}
  \bibinfo{author}{\bibfnamefont{F.}~\bibnamefont{Verstraete}},
  \bibinfo{journal}{Phys. Rev. Lett.} \textbf{\bibinfo{volume}{106}},
  \bibinfo{pages}{170501} (\bibinfo{year}{2011}).

\bibitem[{\citenamefont{Vidal}(2004)}]{VidalPRL04}
\bibinfo{author}{\bibfnamefont{G.}~\bibnamefont{Vidal}},
  \bibinfo{journal}{Phys. Rev. Lett.} \textbf{\bibinfo{volume}{93}},
  \bibinfo{pages}{040502} (\bibinfo{year}{2004}).

\bibitem[{\citenamefont{White and Feiguin}(2004)}]{WhitePRL04}
\bibinfo{author}{\bibfnamefont{S.~R.} \bibnamefont{White}} \bibnamefont{and}
  \bibinfo{author}{\bibfnamefont{A.~E.} \bibnamefont{Feiguin}},
  \bibinfo{journal}{Phys. Rev. Lett.} \textbf{\bibinfo{volume}{93}},
  \bibinfo{pages}{076401} (\bibinfo{year}{2004}).

\bibitem[{\citenamefont{Daley et~al.}(2004)\citenamefont{Daley, Kollath,
  Schollw\"ock, and Vidal}}]{DaleyJSM04}
\bibinfo{author}{\bibfnamefont{A.~J.} \bibnamefont{Daley}},
  \bibinfo{author}{\bibfnamefont{C.}~\bibnamefont{Kollath}},
  \bibinfo{author}{\bibfnamefont{U.}~\bibnamefont{Schollw\"ock}},
  \bibnamefont{and} \bibinfo{author}{\bibfnamefont{G.}~\bibnamefont{Vidal}},
  \bibinfo{journal}{J. Stat. Mech.: Theory Exp.}
  \textbf{\bibinfo{volume}{2004}}, \bibinfo{pages}{P04005}
  (\bibinfo{year}{2004}).

\bibitem[{\citenamefont{Poulin}(2010)}]{PoulinPRL10}
\bibinfo{author}{\bibfnamefont{D.}~\bibnamefont{Poulin}},
  \bibinfo{journal}{Phys. Rev. Lett.} \textbf{\bibinfo{volume}{104}},
  \bibinfo{pages}{190401} (\bibinfo{year}{2010}).

\bibitem[{\citenamefont{Hastings}(2004)}]{HastingsPRL04}
\bibinfo{author}{\bibfnamefont{M.~B.} \bibnamefont{Hastings}},
  \bibinfo{journal}{Phys. Rev. Lett.} \textbf{\bibinfo{volume}{93}},
  \bibinfo{pages}{140402} (\bibinfo{year}{2004}).

\end{thebibliography}

\begin{thebibliography}{11}
\expandafter\ifx\csname natexlab\endcsname\relax\def\natexlab#1{#1}\fi
\expandafter\ifx\csname bibnamefont\endcsname\relax
  \def\bibnamefont#1{#1}\fi
\expandafter\ifx\csname bibfnamefont\endcsname\relax
  \def\bibfnamefont#1{#1}\fi
\expandafter\ifx\csname citenamefont\endcsname\relax
  \def\citenamefont#1{#1}\fi
\expandafter\ifx\csname url\endcsname\relax
  \def\url#1{\texttt{#1}}\fi
\expandafter\ifx\csname urlprefix\endcsname\relax\def\urlprefix{URL }\fi
\providecommand{\bibinfo}[2]{#2}
\providecommand{\eprint}[2][]{\url{#2}}







\bibitem[{\citenamefont{Kubo}(1957)}]{Kubo57}
\bibinfo{author}{\bibfnamefont{R.}~\bibnamefont{Kubo}}, \bibinfo{journal}{J.
  Phys. Soc. Jpn.} \textbf{\bibinfo{volume}{12}}, \bibinfo{pages}{570}
  (\bibinfo{year}{1957}).



\bibitem[{\citenamefont{Suzuki}(1994)}]{SuzukiCMP94}
\bibinfo{author}{\bibfnamefont{M.}~\bibnamefont{Suzuki}},
  \bibinfo{journal}{Commun. Math. Phys.} \textbf{\bibinfo{volume}{163}},
  \bibinfo{pages}{491} (\bibinfo{year}{1994}).

\bibitem[{\citenamefont{Rodriguez}(2009)}]{RodriguezAML09}
\bibinfo{author}{\bibfnamefont{J.~A.} \bibnamefont{Rodriguez}},
  \bibinfo{journal}{Appl. Math. Lett.} \textbf{\bibinfo{volume}{22}},
  \bibinfo{pages}{916} (\bibinfo{year}{2009}).

\bibitem[{\citenamefont{Dawson and Nielsen}(2006)}]{DawsonQIC06}
\bibinfo{author}{\bibfnamefont{C.~M.} \bibnamefont{Dawson}} \bibnamefont{and}
  \bibinfo{author}{\bibfnamefont{M.~A.} \bibnamefont{Nielsen}},
  \bibinfo{journal}{Quant. Inf. Comput.} \textbf{\bibinfo{volume}{6}}
  (\bibinfo{year}{2006}).

\end{thebibliography}
\end{document}